\documentclass[12pt,modern]{aastex62} 
\usepackage{verbatim}
\usepackage{natbib,aas_macros}
\usepackage{hyperref}
\usepackage{graphicx}
\usepackage{xspace}
\usepackage{xcolor}  
\usepackage{rotating} 

\newcommand{\cchp}{\,CCHP\xspace}                        
\newcommand{\cspi}{\,CSP-I\xspace}                       
\newcommand{\gaia}{\emph{Gaia}\xspace}            
\newcommand{\ho}{$H_{0}$\xspace}                  
\newcommand{\hounits}{\,km\,s$^{-1}$\,Mpc$^{-1}$\xspace}   
\newcommand{\hst}{\emph{HST}\xspace}              
\newcommand{\hstwfciii}{\emph{HST/WFC3}\xspace}    
\newcommand{\hstacs}{\emph{HST/ACS}\xspace}   
\newcommand{\jwst}{\emph{JWST}\xspace}  
         
\newcommand{\lcdm}{$\Lambda$CDM\xspace}

\newcommand{\sne}{SNe~Ia\xspace}                   
\newcommand{\sn}{SN~Ia\xspace}                   
\newcommand{\sig}{$\sigma$\xspace}               
\newcommand{\shoes}{\emph{SHoES}\xspace}            
\newcommand{\ngc}{NGC\,\xspace}
\newcommand{\ic}{IC\,1613\xspace}

\defcitealias{beaton_2016}{Paper I}
\defcitealias{hatt_2017}{Paper II}
\defcitealias{jang_2018}{Paper III}
\defcitealias{hatt_2018a}{Paper IV}
\defcitealias{hatt_2018b}{Paper V}
\defcitealias{hoyt_2019}{Paper VI}
\defcitealias{beaton_2019}{Paper VII}

\shorttitle{The Carnegie-Chicago Hubble Program}
\shortauthors{Freedman et al.}

\begin{document}
\title{\textit{The Carnegie-Chicago Hubble Program}. VIII. An Independent Determination of the Hubble Constant Based on the Tip of the Red Giant Branch\protect\footnote{Based on observations made with the NASA/ESA Hubble Space Telescope, obtained at the Space Telescope Science Institute, which is operated by the  Association of Universities for Research in Astronomy, Inc., under NASA contract NAS 5-26555. These observations are associated with programs \#13472 and \#13691.}}

\bigskip

\author[0000-0003-3431-9135]{Wendy~L.~Freedman}\affil{Department of Astronomy \& Astrophysics, University of Chicago, 5640 South Ellis Avenue, Chicago, IL 60637, USA}
\email{wfreedman@uchicago.edu}
  
\author{Barry~F.~Madore}
\affil{The Observatories of the Carnegie Institution for Science, 813 Santa Barbara St., Pasadena, CA 91101, USA}

\author[0000-0003-2767-2379]{Dylan~Hatt}
\affil{Department of Astronomy \& Astrophysics, University of Chicago, 5640 South Ellis Avenue, Chicago, IL 60637, USA}

\author{Taylor~J.~Hoyt}
\affil{Department of Astronomy \& Astrophysics, University of Chicago, 5640 South Ellis Avenue, Chicago, IL 60637, USA}

\author{In~Sung~Jang}  
\affil{Leibniz-Institut f\"{u}r Astrophysik Potsdam, D-14482 Potsdam, Germany}
    
\author[0000-0002-1691-8217]{Rachael~L.~Beaton}    
\affil{Department of Astrophysical Sciences, Princeton University, 4 Ivy Lane, Princeton, NJ~08544, USA}

\author{Christopher~R.~Burns}
\affiliation{Observatories of the Carnegie Institution for Science, 813 Santa Barbara St, Pasadena, CA, 91101, USA}

\author{ Myung~Gyoon~Lee}
\affil{Department of Physics \& Astronomy, Seoul National University, Gwanak-gu, Seoul 151-742, Korea}

\author{ Andrew~J.~Monson}
\affil{Department of Astronomy \& Astrophysics, Pennsylvania State University, 525 Davey Lab, University Park, PA 16802, USA}

\author{Jillian~R. Neeley}
\affil{Department of Physics, Florida Atlantic University, 777 Glades Rd, Boca Raton, FL 33431}

\author{M.~M.~Phillips}
\affiliation{Carnegie Institution of Washington, Las Campanas Observatory, Casilla 601, Chile}
 
\author{Jeffrey~A.~Rich}
\affil{Observatories of the Carnegie Institution for Science, 813 Santa Barbara St., Pasadena, CA~91101, USA}

\author[0000-0002-1143-5515]{Mark~Seibert}
\affil{Observatories of the Carnegie Institution for Science, 813 Santa Barbara St., Pasadena, CA~91101, USA}

\begin{abstract}

We present a new and independent determination of the local value of the Hubble constant based on a calibration of the Tip of the Red Giant Branch (TRGB) applied to Type Ia supernovae (\sne). We find a value of \ho = 69.8 $\pm$ 0.8 ($\pm$1.1\% stat) $\pm$ 1.7 ($\pm$2.4\% sys) \hounits.  The TRGB method is both precise and accurate, and is parallel to, but independent of the Cepheid distance scale. Our value sits midway in the range defined by the current Hubble tension. It agrees at the 1.2\sig level with that of the \citet{planck_2018} estimate, and at the 1.7\sig level with the \hst \shoes  measurement of \ho  based on the Cepheid distance scale.  The TRGB distances have been measured using deep {\em Hubble Space Telescope (HST) Advanced Camera for Surveys (ACS)} imaging of galaxy halos.  The zero point of the TRGB calibration is set with a distance modulus to the Large Magellanic Cloud of 18.477 $\pm $ 0.004 (stat) $\pm $0.020 (sys) mag, based on measurement of 20 late-type detached eclipsing binary (DEB) stars, combined with an \hst parallax calibration of a 3.6 $\mu$m Cepheid Leavitt law based on {\it Spitzer} observations. We anchor the TRGB distances to galaxies that extend our measurement into the Hubble flow using the recently completed Carnegie Supernova Project I (\cspi) sample containing about 100 well-observed \sne. There are several advantages of halo TRGB distance measurements relative to Cepheid variables: these include low halo reddening, minimal effects of crowding or blending of the photometry, only a shallow (calibrated) sensitivity to metallicity in the $I$-band, and no need for multiple epochs of observations or concerns of different slopes with period. In addition, the host masses of our TRGB host-galaxy sample are higher on average than the Cepheid sample, better matching  the range of host-galaxy masses in the \cspi distant sample, and reducing potential systematic effects in the \sne measurements.

\end{abstract}


\section{Introduction}
\label{sec:intro}

The Hubble constant (\ho), which parameterizes the current expansion rate of the universe, plays a critical role in cosmology by setting the absolute size scale for the universe.  In recent decades, remarkable progress has been made in improving the accuracy (by identifying and decreasing the systematic errors) in measurements of \ho. From a factor-of-two uncertainty in measuring extragalactic distances only a few decades ago, a value of \ho measured to 10$\%$ was made possible with the availability of \hst \citep{freedman_2001}; and more recently the uncertainties have been reduced to less than 5\% by a number of investigations \citep[e.g.,][]{freedman_madore_2010, freedman_2012, riess_2016,  suyu_2017}.

Currently a  diverse set of  increasingly precise measurements have led to convergence on a standard cosmology: a model of the universe whose energy + matter density is dominated by dark energy (in the form of a cosmological constant, $\Lambda$)  and cold dark matter (CDM). This concordance \lcdm model is consistent with a wide array of independent observations including, but not limited to,  measurement of anisotropies in the temperature and polarization of the cosmic microwave background (CMB) \citep[e.g.,][]{bennett_2013, planck_2018};
fluctuations in the density of baryonic matter or baryonic acoustic oscillations (BAO), \citep[e.g.,][]{eisenstein_2005, cole_2005, aubourg_2015, alam_2017}; and observations of the magnitude-redshift relation for high redshift  \sne, \citep[e.g.,][]{riess_1998, perlmutter_1999, betoule_2014, scolnic_2018}. 

The temperature and polarization anisotropy spectra for the \citet{planck_2018} data are extremely well fit by a 6-parameter-only \lcdm model. While some parameters are derived from the CMB measurements with extremely high precision (e.g., the angular size of the sound horizon, which is measured to an extraordinary precision of $\pm$0.03\%, \citep{planck_2018}), the CMB measurements themselves do not give a {\it direct} measure of \ho. They provide instead an {\it indirect} constraint -- with a very small uncertainty -- {\it but only under the assumption of this 6-parameter cosmological model. Assuming  this standard \lcdm model}, the Planck Collaboration infers a value of the Hubble constant of 67.4 $\pm$ 0.5  \hounits.

This CMB-modeled value of \ho stands in stark contrast with two decades of (systematically larger) determinations of the local value of \ho \citep[e.g.,][]{freedman_2001, riess_2011, freedman_2012, sorce_2013, riess_2016,  suyu_2017, birrer_2018, burns_2018}. 
The determination of  \citet{riess_2019}, \ho = 74.03 $\pm$ 1.42 \hounits,  is $+6.6$ \hounits larger and 4.4\sig discrepant with the above value  of 67.4 $\pm$ 0.5  \hounits quoted by \citet{planck_2018}.
It represents a 10\% difference between the two distance scales. Moreover, the significance of this divergence has been increasing with time. While the locally-determined value of \ho has not changed appreciably (the value of the Hubble constant determined by the HST Key Project nearly 20 years ago was 72 $\pm$ 8 \hounits), both the precision and the accuracy of the local measurements have increased considerably, with the quoted errors dropping from 10\% in 2001 to less than 3\% in 2018. The Planck errors have been consistently smaller and stable, at the 2\% level when they were first reported in 2014, and now $<$1\% in 2018.

An additional means of determining \ho 
is based on measurements of fluctuations in the matter density resulting from the imprint of BAO at recombination \citep{aubourg_2015,alam_2017,macaulay_2018}.   Applying an `inverse' distance ladder approach, the {\it absolute}  calibration for the distance scale (or equivalently, the absolute magnitude calibration of \sne) is set by adopting a sound horizon scale for the CMB (and not from Cepheids or the TRGB, as in the local universe measurements). The sound horizon, $r_s$, depends on early-time physics, requiring knowledge of the  density  and  equation  of  state  parameters  of  different  species  in  the  early universe. Adopting a sound horizon scale of $r_s$ = 147.05 $\pm$ 0.30 Mpc ($\pm$0.2\%) \citep{planck_2018},  Macaulay et al. most recently obtain a value of \ho = 67.77 $\pm$ 1.30 \hounits, based on a sample of 329 \sne and BAO measurements from the Dark Energy Survey (DES). In this analysis, the {\it relative} \sne distance measurements serve to extrapolate the (also relative) BAO measurements made at larger redshifts down to redshift z=0.  Therefore, the BAO calibration of \ho is not completely independent of the Planck measurement, since both \ho determinations are based on the standard $\Lambda$CDM model and its adopted value of the sound horizon scale.  However, the combination of BAO,  CMB, and \sne measurements provides a very powerful constraint on the shape of the distance-redshift relation from z=1100 to the  present day, limiting possible variations in H(z) that might potentially alleviate some of the current tension in \ho.

\subsection{Possible Sources of the Tension in H$_0$}

Given the increasing tension with the Planck results over time, it is critically important to enumerate and assess quantitatively the impact of all systematic uncertainties that may still be affecting one (or more) of these methods and models \citep{freedman_2017}.  The stakes for resolving this tension are particularly high:  the persistent tension may be signaling fundamental physics beyond the baseline \lcdm standard model. 

Many possible explanations for the $6$ \hounits discrepancy have been considered in the recent literature; but to date, all have been found lacking. Two long-standing questions have been discussed extensively and resolved: (1) the issue of whether we live in a local bubble, and (2) the effects of weak lensing on the \sne measurements.

The question of whether or not we live in an underdense local void or bubble has recently been re-addressed by \citet{wu_huterer_2017,hoscheit_barger_2018,darcy_kenworthy_2019} (and see references therein). They conclude, in agreement with previous studies, that an effect of this kind is too small to explain the magnitude of observed tension. In their detailed numerical simulations, Wu \& Huterer find that the typical sample variance in \ho in the local universe amounts to $\pm$0.31 \hounits, more than an order of magnitude smaller than the observed discrepancy, and they further conclude that the existence of a  void of the required size  is of negligible probability in a $\Lambda$CDM model.

The second long-standing question is whether an effect due to  weak lensing could be systematically affecting the dispersion of \sne magnitudes as distant \sne are lensed by matter along the line of sight \citep[e.g.][]{frieman_1996, holz_1998}.
\sne will be magnified in brightness when the lensing convergence is positive, and then de-magnified when it is negative. The effect will increase with increasing redshift, as  longer path-lengths are traversed.  Once more, however, this effect is found to be far too small to explain the measured \ho difference  \citep[see, for example,][and references therein]{smith_2014}. While the expected lensing effects are seen to modestly increase the scatter in the observed Hubble diagram, they do not contribute in a systematic way to the measurement of \ho.

If the tension persists, and cannot be attributed to a known astrophysical effect or systematic error, what are the alternatives? The observed tension could be signaling additional fundamental new physics, either beyond  the current astronomers$'$ 6-parameter standard \lcdm model, or beyond the physicists$'$  standard model of particle physics.  At present, the dominant components of the standard model of cosmology are dark energy and dark matter, neither of which has a firm theoretical foundation. Simple examples of physics beyond the standard \lcdm model could include  evolution of the dark energy equation of state, or an increase in the energy density of radiation in the early universe, which would modify the early expansion history of the universe \citep[e.g.,][]{bernal_verde_riess_2016,mortsell_dhawan_2018}. At present, however, these types of {\it late-time} modifications do not lead to a clear improvement in the cosmological model \citep[see, for example][]{planck_2015,  planck_2018}, and references therein.

Indeed, the simplest proposed changes (for example, increasing the number of relativistic species or changing the equation of state for dark energy) worsen the model fit to the CMB anisotropy spectra, and lead to conflicts with measurements from BAO, weak lensing, \sne and/or Big Bang Nucleosynthesis (BBN) measurements  \citep{planck_2018}. More recently, there has been renewed attention to exploring {\it early-time} additional new physics operating prior to recombination \citep[e.g.,][]{karwal_kamionkowski_2016, poulin_2018, agrawal_2019}. These early-universe models  can provide a fit to the measured CMB spectra, although some fine tuning is required to preserve the CMB fits at late times.  Alternatively, non-Gaussian primordial fluctuations in the CMB resulting from long-wavelength (super-CMB) modes may offer a means of explaining some of the observed tension \citep{adhikari_huterer_2019}.  It remains the case that resolving a $+6$ \hounits difference in \ho presents a considerable challenge for theory. Even with these non-trivial theoretical challenges, a reasonable general question remains: do we yet have a complete cosmological model?   The issue remains open at present. With a nod to Carl Sagan, claims for exotic new physics beyond the standard model demand independent and `extraordinary evidence'.

\subsection{Improving the Local Measurements of H$_0$}

The strongest evidence at present for a high value of \ho ($>$ 70 \hounits) rests on an empirical Cepheid calibration of the distances to galaxies hosting \sne (e.g., Riess et al. 2016, 2019; hereafter, \shoes). Several re-analyses of the earlier  \shoes data \citep[e.g.,][]{efstathiou_2014, zhang_2017, follin_knox_2018, feeney_2018}  find statistical consistency with the original analysis of \shoes; however, in all of these cases, the starting point is the same set of reduced data for the Cepheids, as previously analyzed and published as part of the \shoes program.\footnote{\citet{efstathiou_2014} undertook a re-analysis of the \citet{riess_2011} data,  finding somewhat lower values of \ho = 70.6 $\pm$ 3.3 and 72.5 $\pm$ 2.5  \hounits, depending on what local calibration he adopted.} If, for example, there are as-yet-unrecognized systematic errors in the published Cepheid photometry, all follow-up studies would be blind to them. An accurate determination of \ho rests squarely on an accurate determination of the zero point of the extragalactic distance scale. This situation argues strongly for having an alternative method that is completely independent of the Cepheids, capable of providing its own absolute calibration for local measurements of \ho.

The Carnegie-Chicago Hubble Program (\cchp) has been specifically designed to provide this alternative route to the calibration of \sne, and thereby provide an independent determination of \ho via measurement of the Tip of the Red Giant Branch (TRGB) in nearby galaxies.\footnote{In the Carnegie Hubble Program (CHP), \citet{freedman_2012} based their value of \ho on the Cepheid distance scale.} This method has a precision equal to or better than the Cepheid period-luminosity relation (the Leavitt Law) and its current accuracy is also comparable. In five recent papers, we have presented TRGB distances to nine galaxies that are host to 11 \sn, discussed the calibration of the data, and undertaken extensive artificial star tests and error analyses: NGC 1365 \citep[][hereafter, Paper  III]{jang_2018}; \ngc 4424, NGC 4526, NGC 4536 \citep[][hereafter, Paper IV]{hatt_2018a}; \ngc 1316, \ngc 1448: \citep[][hereafter, Paper V]{hatt_2018b};
M66, M96: \citep[][hereafter, Paper VI]{hoyt_2019};
and M101: \citep[][hereafter, Paper VII]{beaton_2019}.
We have also undertaken a detailed comparison of the Cepheid, RR Lyrae and TRGB distances within the nearby Local Group galaxy, \ic \citep{hatt_2017} (Paper II). An overview of the observing program has been presented by \citet{beaton_2016} (Paper I). In addition, we have now begun an extension of the calibration of the TRGB to near-infrared ($JHK$) wavelengths in two nearby galaxies: \ic \citep{madore_2018} and the LMC \citep{hoyt_2018}, which will provide added advantages for future studies, both from the ground, and especially in space.

In this paper, we provide a summary of progress to date on the \cchp, and apply the optical $I$-band TRGB calibration to the third release of \sn data from the Carnegie Supernova Project I (\cspi). An overview of the  \cspi is given  in  \citet{hamuy_2006}. The \cspi has provided a well-observed, multi-wavelength sample of \sne recently published by \citet{krisciunas_2017}.    The outline of the paper is as follows.  In \S \ref{sec:description} we discuss in more detail the motivation for the CCHP and describe the target galaxies and observations. In \S\ref{sec:TRGB}
we review the TRGB method, its calibration and its application to our targets; and then provide a summary of the uncertainties in the method. In \S\ref{sec:trgbceph} we compare the distances measured using the TRGB to those obtained for the same galaxies using Cepheids.
In \S\ref{sec:nearbyHo} we discuss the very nearby (out to distances of 30 Mpc) Hubble diagram for the TRGB and Cepheid galaxies.
In \S\ref{sec:supernovae} we  discuss the two independent \sne samples used in this analysis and the calibration for the absolute magnitudes of \sne. In 
\S\ref{sec:Ho} we present the TRGB calibration of \ho, and a comparison with the Cepheid determination.
In \S\ref{sec:others} we place in context other recent \ho determinations and their uncertainties.
In \S\ref{sec:future}  we discuss future prospects for measuring a local value of \ho to higher precision and better accuracy. Finally, in \S\ref{sec:summary} we present a summary and the implications of our program at this juncture.

\section{The Carnegie-Chicago Hubble Project (CCHP)}
\label{sec:description}

\subsection{Overview and Motivation}

As noted previously, the current goal of the \cchp is to increase both the precision and the accuracy of the TRGB method in order to provide an independent calibration of \sne. The steps to a measurement of the local expansion rate through a `distance ladder' are straightforward. Fortunately, in recent years, the  `rungs' in the distance ladder have been both significantly strengthened, and also reduced in number.  Currently only three steps are required for an accurate calibration of the extragalactic distance scale and a determination of \ho: 
\begin{enumerate}
\item An absolute zero-point calibration using geometric techniques (using, for example, trigonometric parallaxes, masers, and/or detached eclipsing binaries);
\item Absolute distances to a sample of galaxies that are hosts to one or more \sn events, which are simultaneously close enough to have their distances measured (using either Cepheids or, in the case of this paper,  TRGB stars); and
\item High-precision relative distances to a statistically significant sample of galaxies far enough into the Hubble flow so that their peculiar velocities are a small fraction of the cosmological recessional velocities (using \sne).

\end{enumerate}

The largest contributors to the  systematic uncertainty in  \ho are the first two items in the above list:   the accuracy of the calibration of the local distance scale (its zero point) and the total number of calibrators available to tie into the more distant Hubble flow. Increasingly larger samples of \sne in the Hubble flow have only a small impact on the total uncertainty in the measurement of  \ho. One of the biggest remaining challenges to the local measurement of \ho is set by the small number of nearby galaxies that are both host to \sne, and that are also within reach of \hst, for measuring Cepheid distances. 
To date, there are only 19 published Cepheid distances for  nearby \sne observed with modern, linear detectors \citep{riess_2016}, resulting from almost 40 years of \sne searches.\footnote{At the time of the Cycle 22 proposal when this study was begun, only 9 Cepheid measurements to \sne were published \citep{riess_2011}.}

\sne are rare events: those in a host galaxy close enough for high-accuracy measurements of Cepheids  with \hst occur, on average, only once every $\sim$2 years. Building up a larger sample with Cepheids alone could take decades. Nor will \jwst help in this regard since Cepheids are relatively blue stars, and \jwst is diffraction limited at 2 $\mu$m. Hence an additional method for measuring distances with similar precision and high accuracy to a larger number of nearby galaxies, is important for a robust measurement of \ho: the TRGB method offers this opportunity. These red stars are also excellent targets for \jwst.

\subsection{Target Galaxies and Description of Observations}

\begin{figure*} 
 \centering
\includegraphics[width=1.0\textwidth]{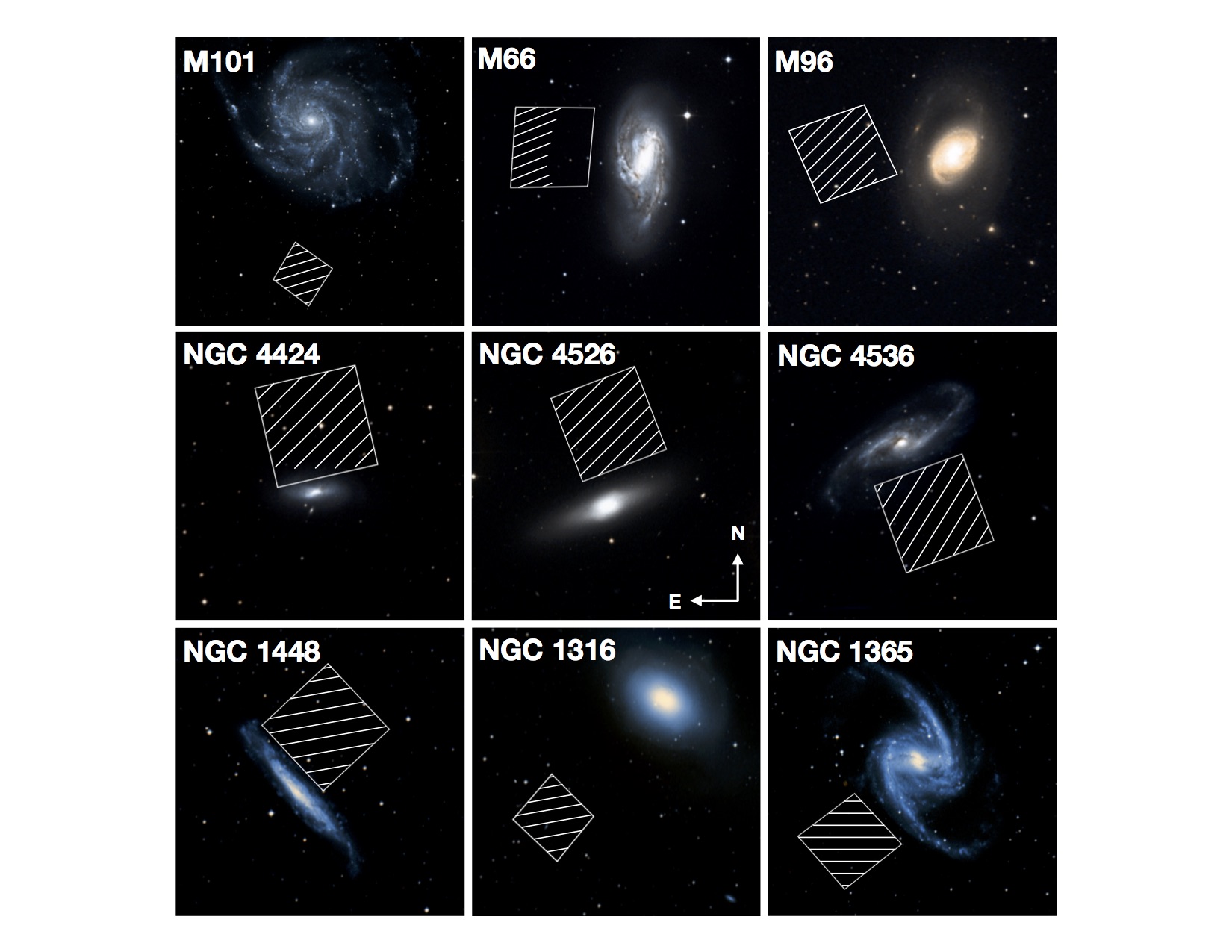}
 \caption{Images of the \hstacs images for the nine TRGB galaxy fields observed as part of our program. The halo target fields in each case are outlined by boxes. The hatched regions indicate those analyzed in this study. North is up, east is to the left.}
 \label{fig:trgbtargets}
\end{figure*}

\begin{figure*} 
 \centering
\includegraphics[width=1.0\textwidth]{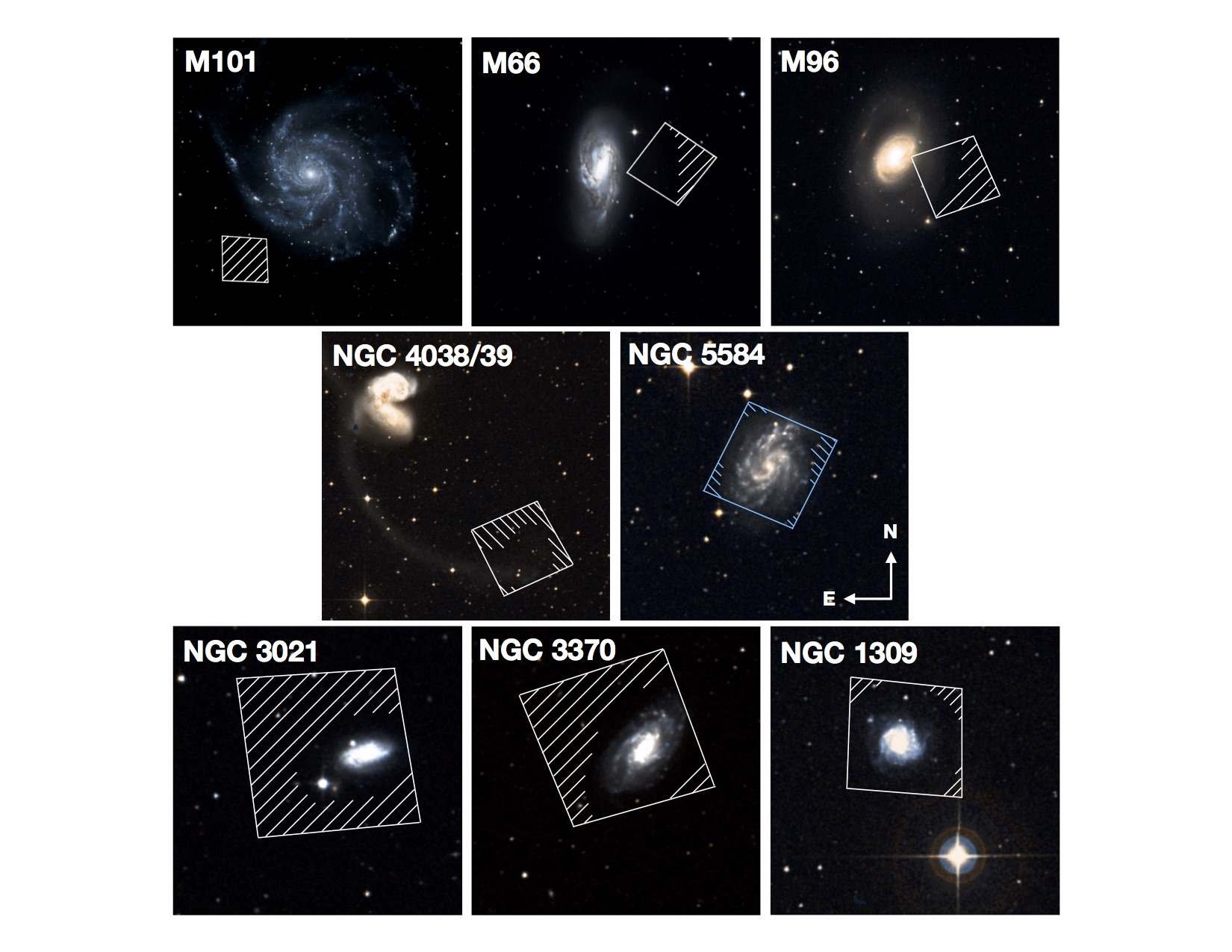}
 \caption{Images of the \hst images for the eight TRGB galaxy fields analyzed by \citet{jang_lee_2017b}. NGC 5584 was observed with $WFC3$; the remaining 7 galaxies were observed with $ACS$. The Jang \& Lee halo target fields in each case are outlined by boxes.The upper three images for M101, M66 and M96 are shown for comparison only; distances to the galaxies have been independently measured as part of the \cchp for the fields illustrated in Figure 1. The hatched regions indicate those analyzed by \citet{jang_lee_2017b}. North is up, east is to the left.}
 \label{fig:trgbtargetsJL}
\end{figure*}

The primary recent focus of the \cchp has been the measurement of the TRGB in the halos of nine galaxies hosting a total of 11 \sne. Observations were taken as part of a \hst Cycle 22 GO proposal \citep[Proposal 13691;][]{hst_2014}. Our targeted galaxies range in distance from 7 Mpc (M101) to almost 20 Mpc (NGC 1316, a member of the Fornax cluster). Data were obtained using the \hstacs and the F814W and F606W filters; total exposure times ranged from 2 $\times$ 1100 sec to 12 $\times$ 1200 sec. The target fields were carefully selected to cover the  halos of the galaxies where the effects of dust are minimal, while simultaneously avoiding contamination by younger and brighter disk AGB stars.  A montage of the CCHP target halo fields is shown in Figure \ref{fig:trgbtargets}. Further details of the observations of the individual galaxies can be found in Papers III-VII. In Figure \ref{fig:trgbtargetsJL}, we show a montage of the fields analyzed by \citet{jang_lee_2017b}.

In Table \ref{tab:trgbgalaxies}, we list the galaxies, their morphological types, foreground $I$-band extinctions, distances (in kpc),  \sn names, and references for the individual distance analyses.
In addition to the 11 \sne calibrators in our sample, we have also re-examined TRGB fits to the archival data for the  galaxies analyzed earlier by \citet{jang_lee_2017b}, yielding an additional 5 galaxies and 5 \sne. Finally, NGC~1404, a galaxy also in the Fornax cluster was host to both SN2007on and SN2011iv. Since we have observations for two Fornax galaxies (NGC~1365 and NGC~1316), we adopt the average distance for these two objects, and apply it to NGC~1404, allowing us to add two further \sn calibrators (SN2007on and SN2011iv) for a total sample of 18.

\begin{deluxetable*}{l l l l l l}
\tablewidth{0pt} 
\tablecaption{TRGB calibration sample \label{tab:trgbgalaxies}} 
\tablehead{\colhead{Galaxy} & \colhead{Type\tablenotemark{
(a)}} & A$_I$ & \colhead{d (kpc)} & \colhead{SN name} & \colhead{Distance Reference} }
\startdata
 &   &    &  \\
\multicolumn{5}{c}{ CCHP Program Galaxies}    \\
&   &    &  \\
    M\,101\tablenotemark{(b)}     & SAB(rs)cd        & 0.013 & 6.5   &  SN 2011fe  & \citet{beaton_2019}; Paper VII \\
    M\,66=NGC 3627\tablenotemark{(b)}      & SAB(S)b          & 0.050 &  11.1 &  SN 1989B  & \citet{hoyt_2019}; Paper VI \\
    M\,96=NGC 3368\tablenotemark{(b)}      & SAB(rs)ab        & 0.038 & 11.6 &   SN 1998bu & \citet{hoyt_2019}; Paper VI  \\
    NGC\,4536  & SAB(rs)bc        & 0.027 & 15.6 &  SN 1981B  & \citet{hatt_2018a}; Paper IV  \\
    NGC\,4526  & SAB0(s),edge-on  & 0.033 & 15.8 &  SN 1994D  & \citet{hatt_2018a}; Paper IV \\
    NGC\,4424  & SAB(s)           & 0.031 & 15.8 &  SN 2012cg & \citet{hatt_2018a}; Paper IV  \\
    NGC\,1448  & SAcd, edge-on    & 0.021 & 18.4 &  SN 2001el & \citet{hatt_2018b}; Paper V  \\
    NGC\,1365  & SB(s)b           & 0.031 & 18.7 &  SN 2012fr  & \citet{jang_2018}; Paper III \\ 
    NGC\,1316  & SAB(0)peculiar   & 0.031 & 19.6 &  SN 1980N, 1981D,& \citet{hatt_2018b}; Paper V  \\
               &                  &       &   &      2006dd \\
    NGC\,1404  & E1           & 0.017 & 19.3 &  SN 2007on, 2011iv  & this paper \\ 
  \hline
&   &    &  \\
\multicolumn{5}{c}{ Our Adopted Distances for the Jang \& Lee  Galaxies} \\ 
&   &    &  \\
      M\,101\tablenotemark{(b)}     & SAB(rs)cd    & 0.013 & 6.5    &  SN 2011fe &   \citet{jang_lee_2017b} \\
      M\,66\tablenotemark{(b)}      & SAB(S)b      & 0.050 &  11.1 &  SN 1989B  & \citet{jang_lee_2017b} \\
      M\,96\tablenotemark{(b)}      & SAB(rs)ab    & 0.038 & 11.6 &   SN 1998bu & \citet{jang_lee_2017b}  \\
      NGC\,4038  & SB(s)m pec   & 0.070 & 21.7   &  SN 2007sr &   \citet{jang_lee_2017b} \\
      NGC\,5584  & SAB(rs)cd    & 0.059 & 23.1   &  SN 2007af &   \citet{jang_lee_2017b} \\
      NGC\,3021  & SA(rs)bc     & 0.020 & 27.8   &  SN 1995al &   \citet{jang_lee_2017b} \\
      NGC\,3370  & SA(s)c       & 0.046 & 28.5   &  SN 1994ae &   \citet{jang_lee_2017b} \\
      NGC\,1309  & SA(s)bc       & 0.060 & 31.6   &  SN 2002fk &   \citet{jang_lee_2017b} \\      
\enddata
\tablenotetext{a}{From the NASA Extragalactic Database, NED.}
\tablenotetext{b}{New (halo) fields in M\,101, M\,66 and M\,96 were observed as part of the CCHP. Archival fields for these galaxies were analyzed by \citet{jang_lee_2017b}. We adopt the new CCHP distances for these galaxies presented in this paper.}
\end{deluxetable*}

The observed scatter in the magnitude-redshift relation for \sne, observed as part of the \cspi, amounts
to only $\pm$0.10 mag \citep{burns_2018}. 
If this intrinsic scatter applies to the calibrating \sne as well, then for a sample of 18 calibrators, this single uncertainty contributes (0.1/$\sqrt(17)$) = 0.024 mag or 1.1\% to the overall systematic error budget. However, as we shall see in \S\ref{sec:absmag}, the measured dispersion in the absolute magnitudes for the calibrator galaxies is slightly larger than for the distant sample, amounting to $\pm$0.12 mag. This larger scatter reflects the (expected) added uncertainty due to  the fact that these \sne were observed with a variety of telescopes and instruments, not all of which were as well-characterized as the \cspi observations. Moreover, the observed scatter in the calibrators will also include the individual uncertainties in the distances to each of the host galaxies.

\section{The TRGB Distance Scale}
\label{sec:TRGB}
\subsection{Description of the TRGB and its Theoretical Basis}
\label{subsec:TRGB_description}

The TRGB marks the onset of core helium burning (the Helium Flash) for low-mass red giants \citep[see][]{salaris_1997, serenelli_2017}. This feature is observed as a clear discontinuity in the first-ascent red giant branch luminosity function. As such, the TRGB provides a simple, empirically-based  \citep[e.g.,][]{lee_1993,rizzi_2007} feature for measuring distances to nearby galaxies. TRGB stars are distinctively bright and red (M$_I$ = -4.0 mag, $(V-I) \sim$ 1.6 mag)  making them easily identifiable and measurable in the uncrowded halos of {\it all types} of nearby galaxies. From an astrophysical perspective, the theory of giant-branch stellar evolution is a mature and well-understood subject \citep[e.g.,][]{salaris_1997, bildsten_2012, serenelli_2017}. The bolometric TRGB is predicted to be only a weak function of metallicity (for Z $<$ 0.04), and an even weaker function of mass (for evolved stars with masses M $<$ 1.4 M$_\odot$ and parent population ages $>$ 4 Gyr).

All of these (empirical and theoretical) characteristics combine to make the TRGB a superb standard candle. To date, most of the TRGB distance determinations have been carried out using the $I$-band (the F814W filter on \hst), where the bolometric corrections flatten the $I$-band luminosity of the observed TRGB as a function of color/metallicity \citep{salaris_cassisi_1998, cassisi_salaris_2013, serenelli_2017}.

\pagebreak

\subsection{Advantages and Disadvantages of the TRGB Method}

Below we enumerate and describe in more detail the many important advantages that the TRGB method has for measuring distances: 
\begin{enumerate}
\item Most importantly, RGB stars that are located in galaxy halos suffer little reddening/extinction by {\it in situ} dust. 
\item Relative to stars located in the higher surface-density disks of galaxies, halo TRGB stars are quite isolated. As a result, they are minimally affected by crowding/blending effects.
\item Halo populations do not contain a significant population of brighter, intermediate-aged AGB stars that can make an accurate measurement of the TRGB more difficult.

\item The $I$-band TRGB is minimally affected by metallicity. Moreover, the metallicity of a TRGB star manifests itself directly in the star's color \citep{salaris_cassisi_1998}, and thus a spread in metallicity is readily identifiable by a widening of the red giant branch in color. This effect can be corrected for, and has been   calibrated empirically \citep{mager_madore_freedman_2008, madore_2009, rizzi_2007, jang_lee_2017a}. 
\item Because all galaxies contain an old, early generation of stars, the TRGB method can be applied to galaxies of all morphological types, as well as those of all inclinations. Cepheids, in contrast, occur only in late-type (star-forming) galaxies, and are difficult to detect in highly-inclined galaxies.
\item Finally,  from an observational perspective, the TRGB method offers a distinct advantage in observing efficiency. For Cepheids, at least a dozen observations, individually spread over a time baseline of several months, are needed to discover the variables, measure their light curves, and determine their periods, amplitudes and mean magnitudes. Further observations sampling the light curves at one or more additional wavelengths are also needed to correct for reddening and to constrain metallicity effects \citep{freedman_2001,riess_2016}. For the TRGB method  single-epoch exposures made in just two pass-bands are all that are required. The resulting CMD [using $I$ vs ($V-I$) or $I$ vs ($I-J$), say] allows the red giant branch population to be color selected, and the TRGB magnitude is then distinguished by an abrupt discontinuity in the color-selected, marginalized $I$-band luminosity function.

\end{enumerate}

A sometimes-cited disadvantage of the TRGB method with respect to Cepheids is that at optical wavelengths TRGB stars are fainter than most of the Cepheids generally observed in external galaxies, typically those with periods greater than 10 days. However, because 12 phase points are needed to discover Cepheid variables, the total observing time required is actually comparable for both methods.\footnote{This is what allowed \citet[for example, ][]{jang_lee_2017b} to measure the TRGB in distant galaxies targeted by \shoes, by stacking the individual Cepheid frames.} Moreover, the TRGB/Cepheid luminosity ratio reverses at near-infrared wavelengths where  TRGB star luminosities exceed that of 10-day Cepheids. Ironically, for Cepheids in the $H$-band, one of the primary contaminants of these variable stars are these same bright red giant (TRGB) stars that are  both projected onto and located within the disk. 

A cautionary flag for the application of the TRGB method is worth emphasizing: it should not be applied in  high-surface-brightness regions   (e.g., the inner disks or the arms of spiral galaxies). Measurement of the TRGB must be focused on the halos of these galaxies: otherwise, disk asymptotic giant branch (AGB) stars will contaminate the sample and can lead to spurious detections \citep[e.g.,][]{saviane_2004, saviane_2008}, as later pointed out by \citet{schweizer_2008} and independently confirmed by \citet{jang_lee_2015} in the case of the Antennae galaxies, NGC\,4038/39. Further examples of spurious detections are those of Maffei 1 and 2 \citep{wu_2014, tikhonov_2018, anand_2019}.  However, care must still be taken to sufficiently populate the TRGB so that small-number statistics are not an issue \citep[see, for example,][]{madore_freedman_1995, mager_madore_freedman_2008, hatt_2017}. Attention to the selection of appropriate halo TRGB fields is straightforward, however, and these potential problems can be anticipated and largely avoided.

\subsection{Measuring the TRGB}
\label{subsec:TRGB_distances}

The use of RGB stars for measuring distances to nearby objects has had a long history, although only in recent decades has its full potential and utility been demonstrated. Originally \citet{shapley_1918} used bright giants in the color-magnitude diagrams of globular clusters  as one means to gauge the size of the Milky Way; resolution of the brightest giants in M31, M32 and other Local Group galaxies led \citet{baade_1944} to his recognition of two population types. Other historical examples can be found in the review by \citet[][and references therein]{madore_freedman_1999}. In a more modern context \citet{mould_kristian_1986} and \cite{freedman_1988} obtained some of the first CCD observations  of extragalactic stellar populations, and used TRGB stars to measure the distances to several Local Group galaxies. Formalizing the technique further, \citet{lee_1993} introduced a quantitative edge detector for measuring the TRGB, convolving the luminosity function of the giant branch with a Sobel filter of the form [-2, 0, +2]. A Sobel (or gradient) filter determines a discrete first derivative; i.e.,  it is specifically designed for locating  edges or sharp discontinuities. It is widely used for this purpose in image processing and analysis applications \citep[e.g.][]{russ_1992}.

A number of refinements to the basic technique have continued to be explored and applied over the past few decades \citep[e.g.,][]{madore_freedman_1995, sakai_madore_freedman_1996,  cioni_2000, mendez_2002, karachentsev_2003, mager_madore_freedman_2008, karachentsev_2018}. Recently \citet{hatt_2017} and \citet{jang_2018} 
compared results for  six different published variations of the basic edge detector, for \ic and \ngc 1365, respectively. In the case of the nearby galaxy \ic, all six methods yielded agreement in the final distance modulus for the galaxy at the $\sim$0.01 mag level, including  our adopted method described below. In the case of \ngc 1365, \citeauthor{jang_2018} find a measured dispersion of about $\pm$0.04 mag, with some outliers. The method that we have adopted  has been explicitly designed to minimize the uncertainties in the application of the edge detector encountered previously.

As described in detail in \citet{hatt_2017} and applied in Papers III-VII, our adopted procedure is first to smooth the observed giant-branch luminosity function using a non-parametric interpolation technique, (GLOESS: Gaussian-windowed, Locally-Weighted Scatterplot Smoothing).  We introduced  GLOESS smoothing in this context because it is effective at suppressing false (noise-induced) edges, especially in the case of sparsely sampled bins in the luminosity function.  It is a technique we have previously used for fitting variable-star light curves  \citep[see][]{persson_2004,monson_2012, monson_2017}.  After smoothing, we then apply a simple [-1, 0, +1] Sobel edge-detection kernel for the measurement of the TRGB.  

\begin{figure*} 
 \centering
\includegraphics[width=1.0\textwidth]{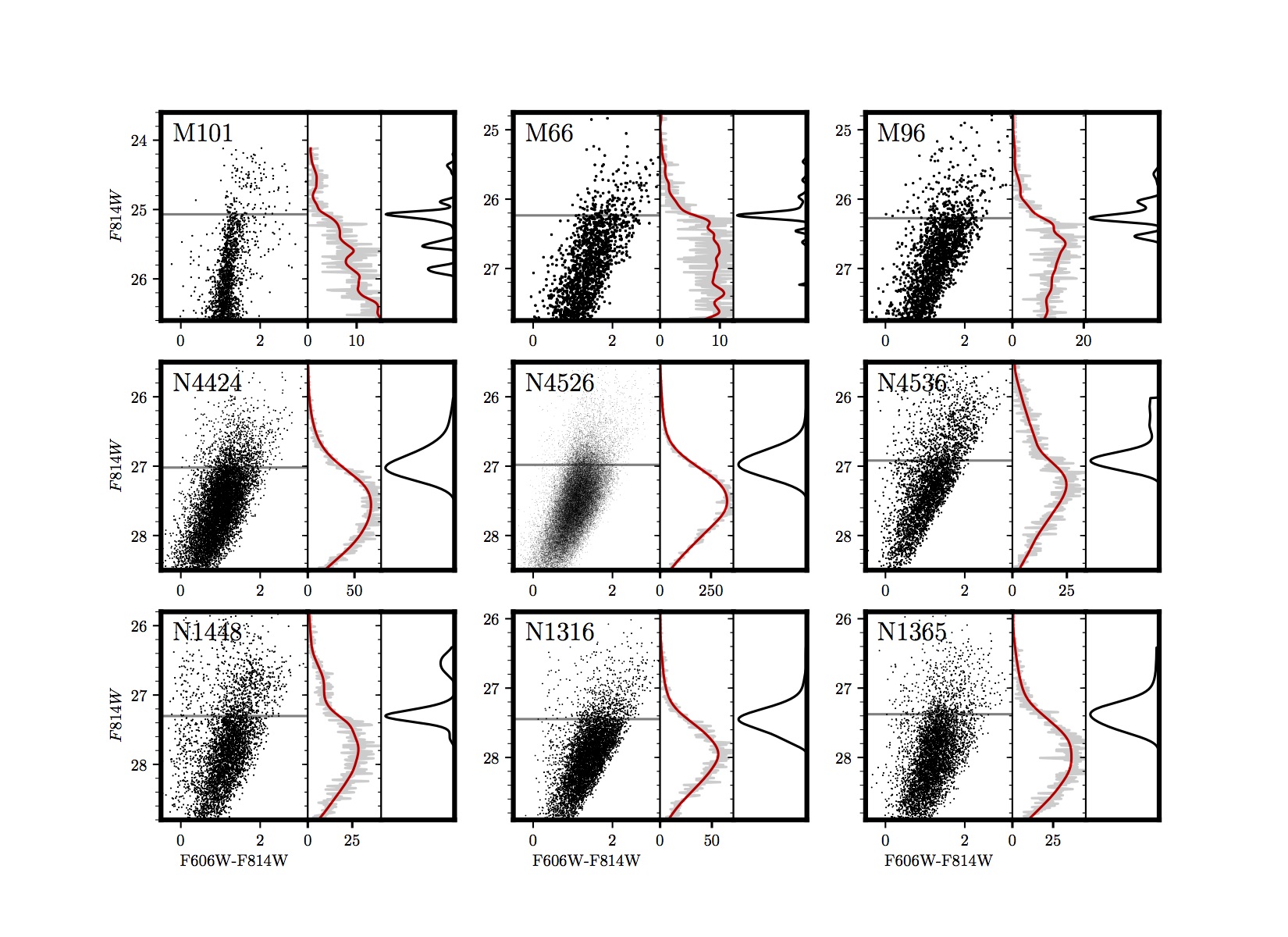}
 \caption{The color-magnitude diagrams (left-hand plots), GLOESS-smoothed luminosity functions (middle plots shown in red) and the edge detection response or Sobel filter output (right-hand plots) for each of the nine \cchp galaxies.  All of the data have been reduced using the same quality cuts based on DAOPHOT sharp and CHI values \citep{stetson_1994}, using the automated pipeline developed for the \cchp \citep[as described  in][]{beaton_2019}. The position of the TRGB in the CMDs is indicated by the black horizontal lines in the CMDs, corresponding to the edge-detection response at the right. }
 \label{fig:cmdlfsobel}
\end{figure*}

For the analysis in this current paper, in order to ensure consistency across all of the galaxies in the program, we have re-reduced all of the \sne host galaxies in the \cchp program with DAOPHOT/ALLFRAME  \citep{stetson_1987,stetson_1994}, using a single, automated pipeline as described in \citet{beaton_2019}. We have constructed a grid of synthetic point spread functions (PSFs) from TinyTim \citep{krist_2011}, and use the same PSF for all of the frames of a given filter for all of the galaxies in the \cchp program.  We have used a set of uniform criteria to define our photometry catalogs across the sample of galaxies, and in some cases, have applied more stringent spatial cuts to avoid contamination by younger AGB stars.  In Figure \ref{fig:cmdlfsobel} we show the updated color-magnitude diagrams, luminosity functions, and Sobel filter output for our nine \cchp targets, used for the analysis in this paper. The edge detection for all of the galaxies was also carried out independently by several of us, and the results only cross-compared in the final stages of analysis.  Next, the GLOESS smoothing was applied in a second pass, stepping iteratively through a range of smoothing scales, to ensure that the peak in the luminosity function was not over-broadened. In order to suppress the effects of  statistical noise fluctuations in any sparsely-sampled bins in the luminosity function,  we weighted the Sobel filter response inversely by the Poisson noise calculated in the adjacent bins in the smoothed luminosity function, as described in \citet{hatt_2017}. The   F814W TRGB magnitudes generally agree to within $\pm$0.04 mag with the values in Papers III-VII. This agreement is reassuring, given that these independent analyses utilize different methods for calculating aperture corrections;  different smoothing scales were used for determining the TRGB magnitudes; and they were carried out by different individuals, in separate analyses intended to allow us to provide  external estimates of the uncertainties.  Overall, our approach is both conceptually simple and, as verified by extensive artificial stars tests \citep{hatt_2017, jang_2018, madore_freedman_2019a} it is also robust, and provides quantifiable uncertainties.

\pagebreak

\subsection{Adopted Zero-Point Calibration for the $I$-band TRGB}
\label{subsec:TRGB_calib}

Our ultimate goal for the absolute calibration of the TRGB is that of geometric parallax measurements for Milky Way RGB stars being obtained by \gaia.\footnote{https://www.cosmos.esa.int/web/gaia} \gaia is a European Space Agency satellite that is measuring parallaxes, proper motions, broad-band photometry, and spectroscopy for over 1 billion stars in the Milky Way, to unprecedented precision. In anticipation of \gaia, we have been using the robotic Three Hundred MilliMeter Telescope (TMMT) at Las Campanas \citep{monson_2017} to obtain accurate $BVI$ photometry for about 2,000 RGB stars in the Milky Way. As discussed in \citet{beaton_2016}, the projected \gaia parallax measurement for these bright giants is expected to ultimately provide a calibration of the TRGB to 0.5\%. 

In the second \gaia release of April 2018 (Data Release, DR2), a significant zero-point offset of $\sim$30$~\mu$as was found with respect to the International Celestial Reference System defined by over 500,000 background quasars \citep{arenou_2018, lindegren_2018}. Given the improvement that will come in future \gaia data releases in 2020 and beyond\footnote{See www.cosmos.esa.int/web/gaia/release}, we have opted in the meantime to anchor our current zero point using  geometric distances to the Large Magellanic Cloud (LMC), as described below. We note that the zero point of the TRGB distance scale remains one of the largest systematic uncertainties in our measurement of \ho. Fortunately, significant further improvement to the \gaia parallaxes will be forthcoming within the next few years.

While awaiting improved parallax results from \gaia, we have updated the absolute-magnitude calibration of the $I$-band TRGB for the LMC. This calibration has four components:
1) the (geometric) distance to the LMC, 2) measurement of the $I$-band TRGB in the LMC, 3) correction for the $I$-band extinction to the LMC TRGB stars, and 4) the transformation from the ground-based Vega system to the \hstacs Vega photometric system. 

\begin{enumerate}

 \item The LMC has two recently determined independent distance moduli, one having a partially-based geometric calibration,  and the other that is directly geometric. \citet{monson_2012} and \citet{scowcroft_2011}  used the {\it Spitzer Space Telescope} to measure mid-infrared 3.6 $\mu$m Period-Luminosity relations (the Leavitt Law)  for  10 Milky Way Cepheids with \hst parallaxes \citep{benedict_2007}
and for a sample of 85 Cepheids in the LMC. Based on these data, \citet{freedman_2012} found a resulting true LMC distance modulus of $\mu_o = $ 18.477 $\pm$ 0.033~mag (sys). A more recent determination that is based on measurements of 20 detached eclipsing binary (DEB) stars in the LMC by \citet{pietrzynski_2019}, gives $\mu_{LMC}$ = 18.477 $\pm$ 0.004 (stat) $\pm$ 0.026 (sys), corresponding to a  distance uncertainty of only 1.2\%.   The DEB value is identical to the \citeauthor{freedman_2012}
modulus, but it has a smaller systematic uncertainty. Combining these two (independent) measurements 
gives our finally-adopted true distance modulus to the LMC of 
$\mu_o = $ 18.477 $\pm$ 0.004 (stat) $\pm$ 0.020~mag (sys) [$\pm$ 1.0\%].

\item We have measured the $I$-band magnitude of the LMC TRGB using the OGLE-III catalog of \citet{ulaczyk_2012}. A Sobel response-function fit to the OGLE-III $I$-band data, excluding an ellipse centered on the bar of the LMC (defined by a = 6.12 deg, b = 1.22 deg, LMC Center: 05 23 34.6 -69 45 22, 
rotation angle = 6 deg), results in a tip detection at  $I$ = 14.595 $\pm$ 0.021 (stat) $\pm$ 0.01 (sys) mag.  Including the entire sample yields exact agreement, indicating that crowding/blending effects are not affecting the result.  We show the color magnitude diagram, luminosity function and Sobel edge-detection filter output in Figure  \ref{fig:ogle_lmc_I_trgb}. 

\begin{figure*} 
 \centering
\includegraphics[width=1.0\textwidth]{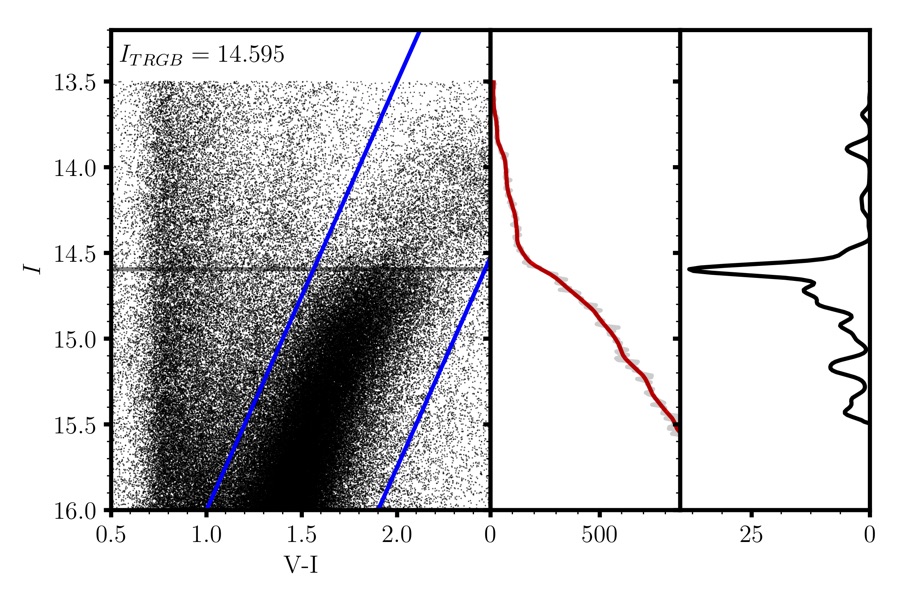}
 \caption{The color-magnitude diagram (left-hand plot), GLOESS-smoothed luminosity function (middle plot shown in red) and the edge detection response or Sobel filter output (right-hand plot) for the outer region of the Large Magellanic Cloud (as defined in the text), based on the OGLE-III catalog from \citet{ulaczyk_2012}. }
 \label{fig:ogle_lmc_I_trgb}
\end{figure*}

\item Recently, \citet{madore_freedman_2019b} have made a direct measurement of the reddening for TRGB stars in the LMC. A brief summary of these results is given in Appendix \ref{App:appendix_lmcred}. The method is similar to that developed for Cepheid variables \citep[e.g.,][]{freedman_1988,freedman_2001, riess_2016}. A comparison of multi-wavelength ($VIJHK$) magnitudes for TRGB stars in the LMC can be made relative to similar multi-wavelength observations of TRGB stars in  other galaxies, and provides a means to measure the {\it differential} extinction for the TRGB stars directly. 
We have compared the multi-wavelength magnitudes for TRGB stars in the LMC [$VI$ OGLE-III data from \citet{ulaczyk_2012}, and $JHK$ cross-matched 2MASS data from \citet{zaritsky_2004}], with those  in  the  SMC \citep{zaritsky_2002} and  in  \ic \citep{hatt_2017}. Correcting these differential measurements for the small, independently-measured SMC and \ic reddenings, then yields the total average reddening to the LMC TRGB stars.  Multi-wavelength differential distance moduli plots for the SMC and \ic can be found in Appendix \ref{App:appendix_lmcred}.  Fits to the multi-wavelength differential moduli result in a determination of the LMC reddening of  $E(B-V) = $ 0.093 $\pm$ 0.01~mag based on the SMC comparison and $E(B-V) = $ 0.096 $\pm$ 0.01~mag based on the IC~1613 comparison. These values are higher than our previously-adopted value of $E((B-V) = $ 0.03 $\pm$ 0.03 mag \citep{hatt_2018a,hatt_2018b, hoyt_2018}; however, the current measurement applies directly to the TRGB stars themselves.  The corresponding extinctions in the I band are $A_I = $ 0.158 $\pm 0.01$ mag and $A_I = $ 0.162 $\pm 0.01$ mag, respectively. Although formally the error is 0.01~mag, we conservatively adopt a mean value of $A_I = $ 0.160 $\pm$ 0.02~mag.\footnote{We have also perturbed the solutions, allowing for a $\pm$0.03 mag uncertainty in the $V$ and $I$ zero points. The average deviation in $A_I$ is only 0.011 mag.} Our mean value of A$_I$ = 0.160 mag, (corresponding to E($V-I$) = 0.131 mag), agrees to within the 1\sig uncertainties (of $\pm$0.02 mag)  with other recent reddening estimates to the LMC, based on a number of different methods \citep[e.g.][]{joshi_2019}. We note that \citet{haschke_2011}  find values of E$(V-I)$ of 0.11 and 0.09 for LMC RR Lyrae stars and red clump stars, respectively, albeit with larger uncertainties of 0.07 and 0.06 mag.

\item To transform from the ground-based $I$-band (Vega-system photometry) of  \citet{ulaczyk_2012} into the Vega-based \hstacs  photometric system in F814W we used the $HST/WFC3$ calibration given by \citet{riess_2016}. We note that the $WFC3$ F814W zero point agrees to within -0.002 mag with that for $ACS$ for F814W\footnote{http://www.stsci.edu/hst/acs/documents/isrs/isr1710.pdf}: 
\begin{equation}
\label{eq:HSTcal}
m_{814W} = I + 0.02 -0.018(V-I) 
\end{equation}
For a fiducial color of ($V-I$) = 1.6 mag for the LMC TRGB, this correction amounts to 0.0088 mag. \citet{riess_2016} concluded that the uncertainty in the zero point of the transformation was $<$0.004 mag. We adopt this uncertainty as the statistical uncertainty alone. We compare this offset with that measured independently by \citet{hatt_2017} for ground-based photometry for standard stars in \ic, transformed to the $ACS$ F814W photometric system, which provides an independent test of the zero point.  \cite{hatt_2017} find that F814W = $I$ - 0.02 (their Figure 11), agreeing to within $\pm$0.01 mag with the value of 0.01 mag from Equation \ref{eq:HSTcal}.\footnote{We note also that the recent recalibration by \citet{riess_2019},  based on observations for LMC Cepheids, results in a difference (in the sense of Ground - \hstwfciii) of  0.018 mag, with a standard deviation of 0.036 mag  for the F814W observations. This offset would result in an absolute magnitude of  $M_I^{TRGB} = -4.067$ mag, (within our adopted uncertainty), which would lead to a 0.9\% decrease in \ho. At this time, we have not applied this (approximate) transformation, which, as \citet{riess_2019} note is for comparison purposes only. Finally, we note that this new calibration, applied to $VIH$ data for the Cepheids has a larger impact on \ho than in the case of the TRGB, resulting from the fact that for the Cepheids, the Wesenheit function, W, as defined by \citet{riess_2019}, is m$_H^W$ = F160W - 0.386 (F555W - F814W). The offsets (Ground - \hst) for $VIH$ and $m_H^W$ after being transformed to on-flight magnitudes, amount to +0.036, +0.018, -0.032, and -0.040 mag, respectively.} Conservatively, we adopt an additional $\pm$0.02 mag systematic uncertainty for the $I$-band to $WFC3$ transformation. We also include an additional $\pm$0.01 mag systematic uncertainty to account for the $WFC3$ to $ACS$  transformation, and an additional $\pm$0.01 mag systematic uncertainty to account for metallicity. 

\pagebreak
Finally, Riess (2019, private communication) has compared the OGLE-III $I$-band photometry with F814W ACS data for 7 short-exposure fields in the LMC, allowing a comparison of red giant branch stars. He finds a median difference of ACS F814W -  $I$(OGLE-III) = 0.0035 mag. This difference agrees well with the difference of F814W - $I$ based on the calibration adopted in Table \ref{tab:lmc_systematics}, which amounts to 0.0088 - 0.002 = 0.0068 mag; i.e., the difference is at the 0.0033 mag level. These various comparisons lend confidence that the current calibration of the LMC TRGB is accurate to within our quoted uncertainties.

\end{enumerate}

When combined with  our currently adopted LMC distance modulus of 18.477 $\pm$ 0.004 (stat) $\pm$ 0.020 (sys) mag, and adding the uncertainties described above in quadrature, this yields $M_I = I_{LMC}^{TRGB} - \mu_{LMC} - A_I - zp = 14.595 - 18.477 - 0.160 -0.0088 +0.002  = $ -4.049~mag $\pm$ 0.022 ~mag [1.0\%] (stat) $\pm$ 0.039 ~mag [1.8\%] (sys) for the {I-}band TRGB  absolute zero-point calibration at the fiducial color of $(V-I) = 1.6$ ~mag. These sources of uncertainty are summarized in Table \ref{tab:lmc_systematics}. Our adopted value of M$_I =  -4.05$ ~mag is consistent with the range that has historically been published, $-4.00 < M^{TRGB}_{I} < -4.05 $ mag \citep[e.g.,][]{lee_1993, bellazzini_2001, rizzi_2007,jang_lee_2017b}, in addition to being consistent with recent \gaia DR2  parallax results for Milky Way TRGB stars given by \citet{mould_clementini_da_costa_2019}.  We note that if we were to force our TRGB calibration to agree with the most recent value of \ho = 74.22 \hounits based on the LMC distance alone \citep{riess_2019}, that would imply a magnitude shift of 0.13 mag, or a value of M$_I^{TRGB}$ = -3.92 mag.  

\begin{deluxetable}{lccc} 
\tabletypesize{\normalsize}
\setlength{\tabcolsep}{0.05in}
\tablecaption{Systematic Uncertainty in the TRGB zero point \label{tab:lmc_systematics}}
\tablewidth{0pt}
\tablehead{ \colhead{Source}  &  \colhead{Value (mag)} & \colhead{$\sigma_{stat}$} & \colhead{$\sigma_{sys}$}  }
\startdata
$I^{LMC}_{TRGB}$ 	          & 14.595  & 0.021 & 0.01 \\
$A^{LMC}_{\mathrm{I}}$	          & 0.160       & ...     &  0.02 \\
Metallicity  &  ... & ... & 0.01 \\
$I$-band to $WFC3$ & 0.0088           &  0.004   & 0.02 \\
$WFC3$ to $ACS$   & -0.002 & ... & 0.01 \\
Adopted LMC true distance modulus      & 18.477 & 0.004 & 0.02 \\
\hline
M$_{814}^{TRGB}$             & -4.049 & 0.022 & 0.039 \\
\hline \hline
\enddata
\end{deluxetable}

\subsection{TRGB Calibrator Distances to \sne Host Galaxies}
\label{subsec:calib_dist}

In Table \ref{tab:trgb_calibrators}, we list the  supernova name, host galaxy name,   followed by the TRGB  distance moduli and uncertainties; the apparent $B^\prime$ peak magnitudes and uncertainties for the \cspi calibrators (see \S\ref{sec:cspi}); the Cepheid distance moduli and uncertainties; the \shoes apparent $B$ peak magnitudes and uncertainties (from 
\citet{riess_2016}, Table 5; and the velocity from NED\footnote{We have compared our adopted velocities with the cosmicflows3 velocities\citep{graziani_2019};  (http://www2.iap.fr/users/lavaux/science/2mrs\_vel.html)
and find, for the 24 galaxies in our sample, that the mean offset between CF-3 and the NED (group-membership-corrected) velocities is 28 km/sec $\pm$ 150 km/sec. However, the choice of velocities for these galaxies is not of importance in the context of measuring \ho; rather the distances alone provide the calibration of \ho .} (corrected for the local flow) for 27 nearby \sne. The TRGB distance moduli  for each of our nine target galaxies were measured as described in \S \ref{subsec:TRGB_distances}, calibrated by adoption of the zero point given in \S \ref{subsec:TRGB_calib} above. Also listed are the galaxies with distances measured and updated by \citet{jang_lee_2017b} (their Table 3), and updated to our adopted TRGB zero point. We have remeasured the TRGB using techniques as similar as possible to those used for the CCHP galaxies, and find excellent agreement, with the exception of the galaxy, NGC 5584. In this case, we have increased the distance modulus uncertainty to $\pm$0.1 mag, to reflect this difference. The Cepheid distances are taken directly from Table 5 of \citet{riess_2016}, which are based on a calibration using Milky Way parallaxes, NGC~4258 masers and  LMC detached eclipsing binaries. 

\begin{deluxetable}{cccccccccccr}
\tabletypesize{\small}
\tablecaption{TRGB Calibrator Distances \label{tab:trgb_calibrators}} 
\tablehead{\colhead{SN} & \colhead{Galaxy} & \colhead{$\mu_{TRGB}$} & \colhead{$\sigma_T$} & \colhead{TRGB} & \colhead{$m_{B^\prime}^{CSP}$} & \colhead{$\sigma_{B^\prime}^{CSP}$} & \colhead{$\mu_{Ceph}$ \tablenotemark{a)}} & \colhead{$\sigma_C$} &
\colhead{$m_B^{SC}$ \tablenotemark{b)}} & \colhead{$\sigma_B^{SC}$} & \colhead{V$_{NED}$\tablenotemark{c)}}
}
\startdata
1980N & N1316 & 31.46 & 0.04 & CCHP & 12.08 & 0.06 & ... & ... & ... & ... & 1306 \\
1981B & N4536 & 30.96 & 0.05 & CCHP & 11.64 & 0.04 & 30.91 & 0.05 & 11.62 & 0.12 & 1050 \\
1981D & N1316 & 31.46 & 0.04 & CCHP & 11.99 & 0.17 & ... & ... & ... & ... & 1306 \\
1989B & N3627 & 30.22 & 0.04 & CCHP & 11.16 & 0.07 & ... & ... & ... & ... & 689 \\
1990N & N4639 & ... & ... & ... & 12.62 & 0.05 & 31.53 & 0.07 & 12.42 & 0.12 & 1050 \\
1994D & N4526 & 31.00 & 0.07 & CCHP & 11.76 & 0.04 & ... & ... & ... & ... & 1050 \\
1994ae & N3370 & 32.27 & 0.05 & JL & 12.94 & 0.05 & 32.07 & 0.05 & 12.92 & 0.12 & 1552 \\
1995al & N3021 & 32.22 & 0.05 & JL & 13.02 & 0.05 & 32.50 & 0.09 & 12.97 & 0.12 & 1886 \\
1998aq & N3982 & ... & ... & ... & 12.46 & 0.05 & 31.74 & 0.07 & 12.24 & 0.12 & 1368 \\
1998bu & N3368 & 30.31 & 0.04 & CCHP & 11.01 & 0.06 & ... & ... & ... & ... & 689 \\
2001el & N1448 & 31.32 & 0.06 & CCHP & 12.30 & 0.04 & 31.31 & 0.04 & 12.20 & 0.12 & 1047 \\
2002fk & N1309 & 32.50 & 0.07 & JL & 13.33 & 0.04 & 32.52 & 0.06 & 13.20 & 0.12 & 1864 \\
2003du & U9391 & ... & ... & ... & 13.47 & 0.09 & 32.92 & 0.06 & 13.47 & 0.11 & 2422 \\
2005cf & N5917 & ... & ... & ... & 12.96 & 0.07 & 32.26 & 0.1 & 13.01 & 0.12 & 2244 \\
2006dd & N1316 & 31.46 & 0.04 & CCHP & 12.38 & 0.03 & ... & ... & ... & ... & 1306 \\
2007af & N5584 & 31.82 & 0.1 & JL & 12.72 & 0.05 & 31.79 & 0.05 & 12.70 & 0.12 & 1983 \\
2007on & N1404 & 31.42 & 0.05 & CCHP & 12.39 & 0.07 & ... & ... & ... & ... & 1306 \\
2007sr & N4038 & 31.68 & 0.05 & JL & 12.30 & 0.15 & 31.29 & 0.11 & 12.24 & 0.11 & 1702 \\
2009ig & N1015 & ... & ... & ... & 13.29 & 0.05 & 32.50 & 0.08 & 13.46 & 0.12 & 2534 \\
2011by & N3972 & ... & ... & ... & 12.63 & 0.05 & 31.59 & 0.07 & 12.49 & 0.12 & 1368 \\
2011fe & M101 & 29.08 & 0.04 & CCHP & 9.82 & 0.03 & 29.14 & 0.04 & 9.75 & 0.12 & 455 \\
2011iv & N1404 & 31.42 & 0.05 & CCHP & 12.03 & 0.06 & ... & ... & ... & ... & 1306 \\
2012cg & N4424 & 31.00 & 0.06 & CCHP & 11.72 & 0.06 & 31.08 & 0.29 & 11.55 & 0.11 & 1050 \\
2012fr & N1365 & 31.36 & 0.05 & CCHP & 12.09 & 0.03 & 31.31 & 0.06 & 11.92 & 0.12 & 1302 \\
2012ht & N3447 & ... & ... & ... & 12.66 & 0.12 & 31.91 & 0.04 & 12.70 & 0.12 & 1447 \\
2013dy & N7250 & ... & ... & ... & 12.23 & 0.07 & 31.50 & 0.08 & 12.31 & 0.12 & 1410 \\
2015F & N2442 & ... & ... & ... & 12.40 & 0.03 & 31.51 & 0.05 & 12.28 & 0.14 & 1271\\
\enddata
\tablenotetext{a)}{~Cepheid distances from Table 5 of \citet{riess_2016}.}
\tablenotetext{b)}{~Supernova peak $B$ magnitudes. \citet{riess_2016} Table 5 gives  (m$_B$ + 5a$_B$). Here 5a$_B$ = 3.5635 has been subtracted to give m$_B$. The definition of a$_B$ is given in Equation 5 of \citet{riess_2016}.}
\tablenotetext{c)}{~Velocities listed here (in km/s) are computed from the galaxy redshifts using the linear multi-attractor model provided by NED.  Some galaxies are also members of groups, and further details of their group membership and adopted velocities  can be found in Appendix \ref{App:appendix_NEDvel}.}

\end{deluxetable}

\subsection{Summary of Uncertainties}
\label{subsec:errbudget}

Here we first provide a description of the known and the potential sources of uncertainty (both statistical and systematic), that have gone into the adopted errors in the measurement of individual TRGB distances as given in columns 3 and 4 of  Table \ref{tab:trgb_calibrators}. We then turn to a discussion of the overall systematic uncertainty for the ensemble sample of TRGB calibrators that is critical for our determination of \ho (in \S\ref{sec:systematics}).

\subsubsection{Uncertainties for Individual TRGB Galaxy Distances}
\label{sec:statistical}
 
\begin{enumerate}
 
 \item {\it Photometric Errors and Edge Detection:} In the previous papers in this series (Papers III-VII), we have provided detailed discussions and tabulations of the errors in the photometry for each of the galaxies in the \cchp sample. These include the photometric errors returned by DAOPHOT, aperture corrections from the PSF magnitudes to a 0.5 arcsec aperture, and correction to infinite radius and transformation to the $ACS$ Vega photometric system. In the case of the LMC, we have transformed our photometry from the ground-based Kron-Cousins $I$-band system to the $ACS$ Vega system, as described in \ref{subsec:TRGB_calib}. The foreground galactic extinction corrections, applied on a galaxy-by-galaxy basis, have been adopted from \citet{schlafly_finkbeiner_2011}, as tabulated in NED. For each of our program galaxies, we have undertaken extensive artificial star tests to quantify the effects of crowding and blending on our photometry, as well as to assess their effects on the edge detection and measurement of the TRGB. In some cases we have opted to be more conservative with the error estimates listed in Table \ref{tab:trgb_calibrators} than in the previously published papers. Overall we impose a minimum error of $\pm$0.04 mag, despite the formal errors, in some cases, being as low as $\pm$0.02 mag.

\item {\it AGB Contribution:} The known presence of oxygen-rich AGB stars in the one-magnitude interval above the TRGB, and at about the same color as the TRGB, acts as a source of elevated baseline noise in the application of edge-detection algorithms. This AGB presence lowers the contrast of the tip discontinuity, increasing the uncertainty in the measurement of the tip location. In practice, the effect of the AGB population on the TRGB is somewhat decreased by the fact that in high-precision data a noticeable drop is seen in the AGB luminosity function in the 0.2~mag interval immediately above the TRGB.  This has the effect of restoring some of the contrast of the TRGB tip discontinuity at the tip for high signal-to-noise data. We explicitly incorporate an AGB population into the derivation of an error budget for the tip detection. As described in detail in \citet{hatt_2017, jang_2018, hatt_2018a, hatt_2018b}, we have quantified this effect in our individual target galaxies using artificial star tests.

\item {\it Metallicity:} As discussed in \S\ref{subsec:TRGB_description},
at optical ($B$ and $V$) wavelengths,  theory predicts and observations confirm that the reddest, high-metallicity stars will exhibit a downward slope of the TRGB  in the color-magnitude diagram \citep[e.g.,][]{mager_madore_freedman_2008,jang_lee_2017b}. However, the bolometric corrections work to flatten the tip in the observed redder $I$-band CMD.  We have found that most of the metal-poor stars observed in our target galaxy halos show a negligible color-magnitude slope, necessitating no correction. At high-metallicity there is a (slight) slope to the $I$-band TRGB; however,  in practice, since the most metal-poor stars are brighter than the metal-rich tip stars, the Sobel edge detector triggers a measurement of the edge discontinuity set by the bluer and brighter, metal-poor stars. The redder (and more metal-rich) stars fall to lower luminosities and thereby no longer  contribute to the detection of the edge in the luminosity function. We have experimented with correcting for the slope of the TRGB as in \cite{mager_madore_freedman_2008}, and find that the results do not differ significantly, in precision or accuracy. We have included a metallicity uncertainty of $\pm$0.01 mag.

\item {\it Absolute Zero Point:} In advance of obtaining a calibration from \gaia parallaxes applied to Milky Way field TRGB stars, we have established our zero point using TRGB stars in the Large Magellanic Cloud.  Following our discussion in \S\ref{subsec:TRGB_calib}, we adopt a systematic uncertainty of  $\pm$0.039 ~mag (1.8\%) for the overall zero point for the $I$-band TRGB calibration.

\end{enumerate}

In summary,  $ \sigma_T $, given for each galaxy in Table \ref{tab:trgb_calibrators},  is defined by $ \sigma_T^2 =  \sigma_{phot}^2 + \sigma_{edge}^2 + \sigma_{Z}^2$ where the photometric error is typically $\pm$0.03 mag, the edge detection error is typically $\pm$0.03 mag, and the metallicity uncertainty is taken to be $\pm$0.01 mag, giving a baseline global statistical TRGB measurement uncertainty of $\pm$0.04 mag. The remaining uncertainties due to potential contamination by AGB stars, are determined for each individual galaxy by the individual artificial star tests.

\subsubsection{Overall Systematic Errors}
\label{sec:systematics}

Most of the sources of systematic uncertainty relevant to the determination of the error for an individual galaxy distance become random uncertainties when collectively applied to the determination of \ho. For example, while an error in the individually-measured aperture corrections will contribute a systematic uncertainty to the distance measurement of a given galaxy, for the ensemble of the calibrating sample, the aperture-correction uncertainties combine simply as random or statistical uncertainties. The individual distance uncertainties do, however, determine the weight that each galaxy distance carries in the calibration of the distant \sne. On the other hand,  two of the sources of uncertainty, those of the LMC ground-based $I$-band transformation to the \hstacs F814W photometric system and the uncertainty on the LMC distance, carry over as systematic uncertainties on the  zero-point calibration, and determination of \ho. We carry these two uncertainties forward in determining the overall systematic error in \ho, and distinguish them from the statistical uncertainties. We return to a discussion of the systematic errors, after application of the TRGB calibration to the distant \sne in \S\ref{sec:Ho}. The final uncertainty in \ho combines, in quadrature, the systematic uncertainty given here, with the uncertainty determined in our   MCMC analysis of the \sne alone, as discussed in \S\ref{sec:trgbcal}.

The largest two  contributing factors to the uncertainty  in the local determination of \ho are: (1) the absolute zero points of the Cepheid and/or TRGB distance scales (as described in \S\ref{sec:statistical}, Point 4) and (2) the numbers of Cepheid and/or TRGB calibrated galaxies hosting SNe Ia \citep{riess_2016, burns_2018}. Regarding the first, improvement to the calibration of the Cepheid and TRGB distance scales will  come from future absolute trigonometric parallax measurements being carried out by Gaia \citep{lindegren_2018}: it is anticipated that direct geometric parallaxes will provide zero points to better than 1\% certainty for both the Cepheid and TRGB distance scales. 

Our study is currently focused on decreasing the second source of systematic uncertainty; that is, by increasing the numbers of TRGB zero-point calibrators for measuring the absolute magnitudes of \sne. The measured scatter in the $B$-band absolute magnitudes of the calibrating sample of \cspi \sne is  $\pm$0.12 mag (see \S\ref{sec:absmag}). With 18 TRGB calibrators,  a simple Frequentist estimate of the uncertainty in the  mean absolute magnitude for \sne amounts to $\pm$0.029 mag (i.e., 0.12 / $\sqrt(17)$ mag), which will contribute a 1.3\% uncertainty to the overall Hubble constant error budget. When the uncertainty in the zero point of the TRGB is resolved by the expected \gaia release (in 2022), the dominant term in the Hubble constant error budget will become the number of TRGB distances calibrating the SNe Ia zero point. 

An additional (but very small) uncertainty is contributed by the scatter in the far-field \sne sample. For a sample of 100 \sne, this contributes $\pm$ 0.10 / $\sqrt (99)$ = $\pm$0.01 mag. In what follows, the  errors for the \cspi \sne are calculated formally from the diagonal elements of the covariance matrix from our MCMC analysis (as discussed below in \S\ref{sec:cspi}).

\section{TRGB and Cepheid Distance Comparison}
\label{sec:trgbceph}

A primary goal in the design of \hst was to optimize the discovery and use of Cepheid variables in the calibration of \ho; and it has proven to be highly effective for this purpose \citep[][and references therein]{freedman_2001, sandage_2006, riess_2016}. 
Indeed, for the past century Cepheids have served as a `gold standard' for measuring the distances to nearby galaxies. The list of strengths for Cepheids in measuring extragalactic distances remains long \citep[e.g., see the reviews by][]{madore_freedman_1991, freedman_madore_2010},   and includes the small observed dispersion in the Leavitt Law, especially at longer wavelengths  \citep[e.g., see the first application by][]{mcgonegal_1982}; their clear signal through their distinctive variability and large amplitudes at short wavelengths; and the ability to correct for, or minimize, systematic effects due to reddening and metallicity using multi-wavelength observations \citep[see the first applications by][]{ freedman_grieve_madore_1985, freedman_1988, freedman_madore_1990}. To date, the distances to 64 nearby spiral and irregular galaxies have been measured using Cepheids (NED: https://ned.ipac.caltech.edu ; 22 June 2018 release). Hence, to be useful as an independent  calibration of \ho, any competing distance indicators must demonstrate that they are quantitatively at least as accurate as the Cepheids.  Unfortunately there are few methods available for which both the statistical and systematic uncertainties rival those of Cepheids. For example, there is only one galaxy, NGC 4258 at 7 Mpc, for which a maser distance has been measured and that  can also be used as a zero-point calibrator \citep{herrnstein_1999,humphreys_2013}. Similarly, an accurate application of the DEB method has been confined to the nearby  LMC \citep{pietrzynski_2013, pietrzynski_2019}.

\begin{figure*} 
 \centering
\includegraphics[width=1.0\textwidth]{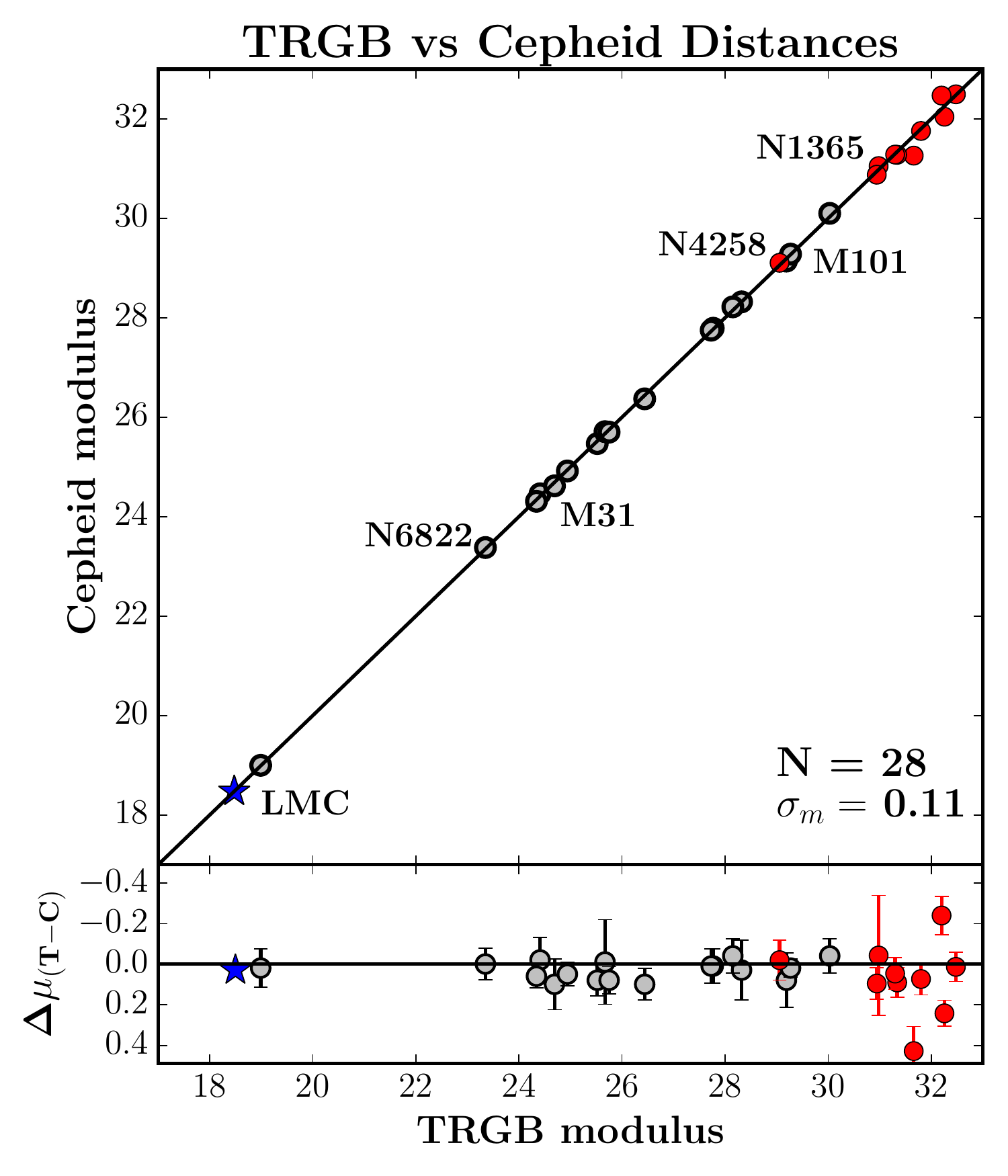}
 \caption{Top panel: A comparison of published TRGB and Cepheid distance moduli for 28 nearby galaxies. The distances span a range from 50 kpc to 30 Mpc. The points in red are the  calibrating galaxies that are host to \sne, used in this study. Gray circled points are from the literature. The LMC, which provides the TRGB zero-point calibration, is shown as a blue star. The black line has a unit slope and the dispersion about the line is $\pm$0.11 mag. Bottom panel: The difference (TRGB minus Cepheid distance modulus) as a function of the TRGB modulus.}
 \label{fig:trgb_ceph_pub}
\end{figure*}

In the upper panel of Figure \ref{fig:trgb_ceph_pub} we show a comparison of published TRGB and Cepheid distance moduli for 28 galaxies. The distances are drawn from the  compilations by \citet{rizzi_2007},  \citet{tully_2013, tully_2015}, \citet{riess_2016} (updated to the Cepheid recalibration of \citet{riess_2019}), NED, and this paper. All of the TRGB distances have been recalibrated to our zero point of M$_I^{TRGB}$ = -4.05 mag. The galaxies span a range of a factor of 60 in distance (50 kpc to 30 Mpc). As can be seen, the overall agreement between the two methods is very good. The $rms$ dispersion about a slope of unity line is $\pm$0.11 mag. A residual difference plot is shown in the lower panel (where $\Delta\mu = \mu_{TRGB} - \mu_{Ceph}$).   The comparison illustrates  the internal consistency of the nearby Cepheid and TRGB distances scales to a level of  5\% in precision in the combined errors, and re-affirms that the TRGB method is competitive with the  Cepheid Leavitt Law \citep[see also earlier studies by][]{lee_1993, ferrarese_2000, rizzi_2007}. Moreover, excluding the 10 distant galaxies that are the hosts to \sne (shown again separately in Figure \ref{fig:trgb_ceph_sngals} below), the dispersion drops to only $\pm$0.05 mag, or 2\% in distance. For the nearest galaxies, the methods show superb agreement.

In Figure \ref{fig:trgb_ceph_sngals}, we show a comparison of TRGB and Cepheid distance moduli focusing on the ten \sne host galaxies that have both TRGB and Cepheid distances, as presented in Table \ref{tab:trgb_calibrators}. Red filled circles represent the galaxies with TRGB distances measured as part of this study. These galaxies cover a range of distances from 7 Mpc (M101) to almost 20 Mpc (NGC 1316).  Red open circles denote the galaxies analyzed by \citet{jang_lee_2017b}. In this case, these galaxies extend out to  distances beyond 30 Mpc. We compare to the  LMC-only calibration for Cepheids from \citet{riess_2019}. The $rms$ scatter about the slope unity line illustrated is $\pm$0.17 mag. The difference ($\Delta\mu = \mu_{TRGB} - \mu_{Ceph}$) is plotted as a function of TRGB distance modulus in the lower panel. The weighted average difference in distance modulus (TRGB-Cepheid) amounts to +0.059 mag.

The scatter in the galaxy-to-galaxy comparison noted above for the TRGB and Cepheid distances amounts to $\pm$0.17 mag. Equally apportioning the resulting uncertainty between the two methods would correspond to an uncertainty of $\pm$0.12 mag for each method, or $\pm$6\% in distance. The scatter seen in Figure \ref{fig:trgb_ceph_sngals} is significantly larger than that for the more nearby galaxies shown in Figure \ref{fig:trgb_ceph_pub} above, where the scatter amounts to only $\pm$0.05 mag, or $\pm$2\% in distance. The scatter in the Cepheid/TRGB overlap sample is also larger than indicated by the individually-measured errors for the individual galaxies, and suggests that the errors in the distances have been underestimated. The source of this additional uncertainty is unknown at present, and it is unclear whether both the TRGB and Cepheid distances share equally in the uncertainties. As we shall see below in \S\ref{sec:nearbyHo}, the scatter in the local Hubble diagram is smaller by a factor of 1.4 for the TRGB-calibrating galaxies than it is for the Cepheid-calibrating galaxies, suggesting that the scatter in the comparison is not equally shared, but rather that the TRGB distances are more precise than  the Cepheid distances.

\begin{figure*} 
 \centering
\includegraphics[width=1.0\textwidth]{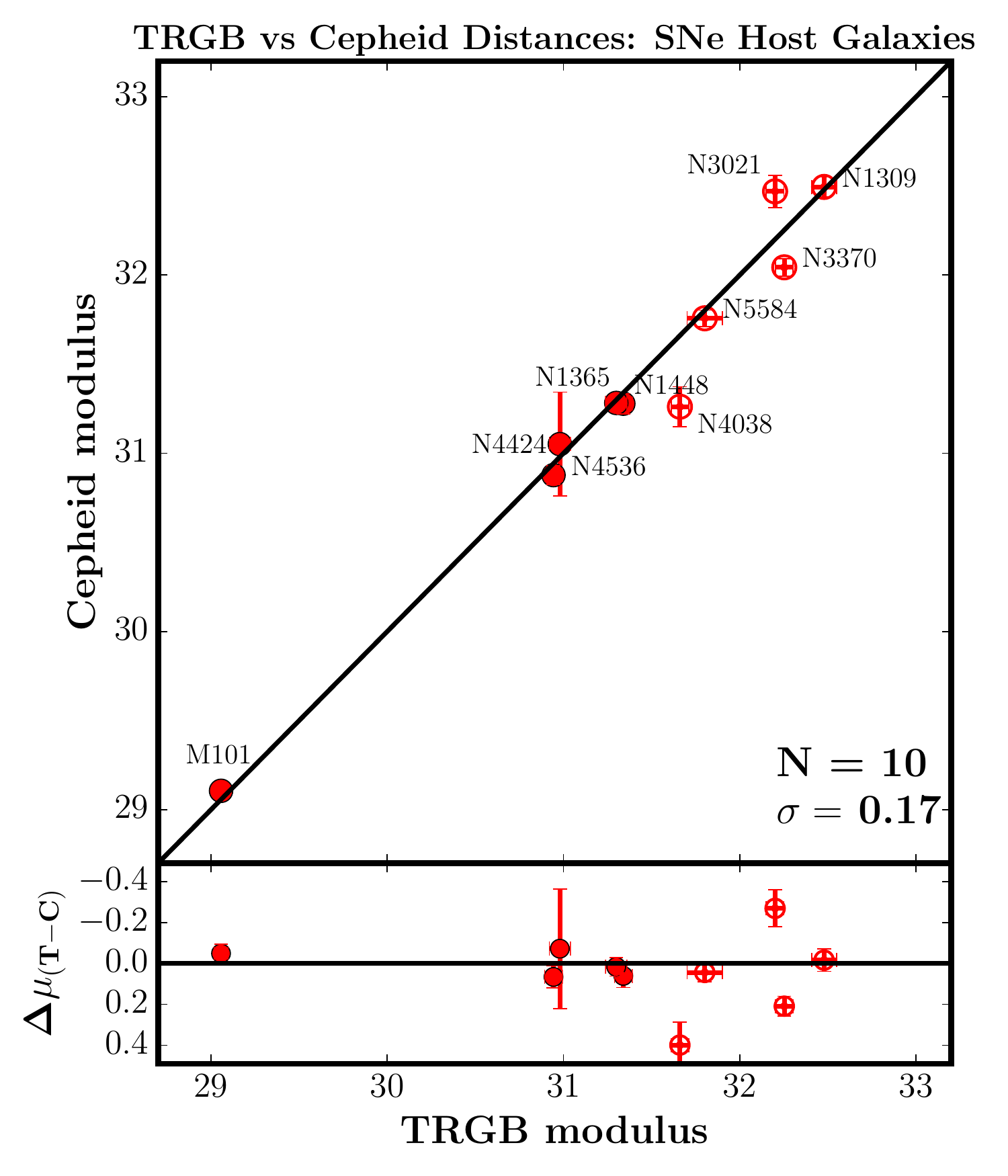}
 \caption{Top panel: An expanded comparison of  TRGB and Cepheid distance moduli for the subset of ten nearby galaxies that are also host to \sne.  The distances span a range from  7 Mpc to over 30 Mpc. Red closed circles are galaxies with distances measured as part of the CCHP. Red open circles are those measured in the study of \citet{jang_lee_2017b}. The black line has a unit slope and the dispersion about the line is $\pm$0.17 mag. Bottom panel: The difference (TRGB minus Cepheid distance modulus) as a function of the TRGB modulus. The weighted average difference in distance modulus (TRGB-Cepheid) amounts to +0.059 mag.} 
 \label{fig:trgb_ceph_sngals}
\end{figure*}

\pagebreak
\section{The Nearby Hubble Diagram}
\label{sec:nearbyHo}

\begin{figure*} 
 \centering
\includegraphics[width=1.0\textwidth]{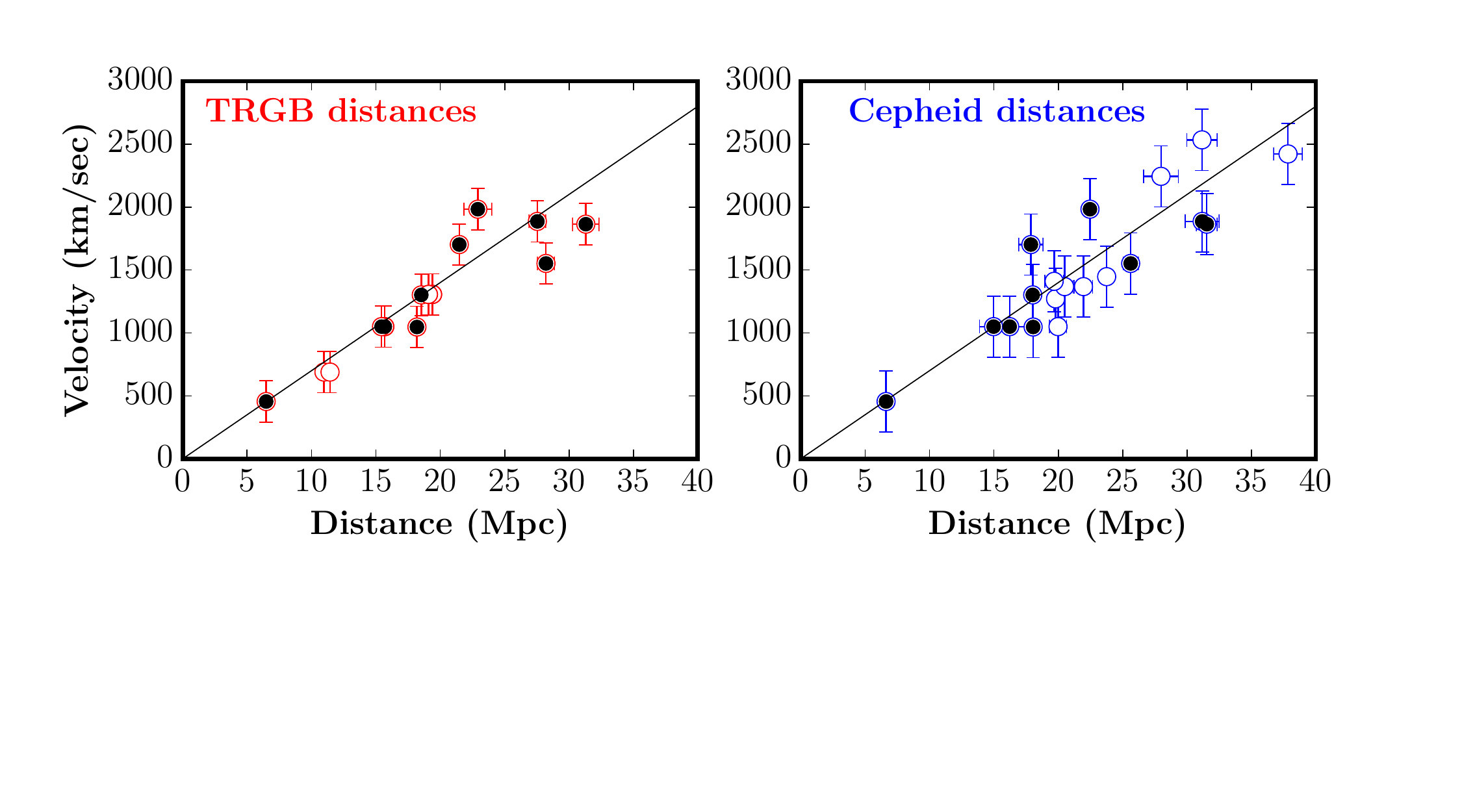}
 \caption{The nearby Hubble diagram for galaxies with TRGB (left panel) and Cepheid (right panel) distances. The filled black circles are for galaxies in common to the TRGB and Cepheid samples. As described in the text, the velocities to these objects have been corrected for the presence of nearby mass concentrations using NED, as described in \S \ref{sec:nearbyHo} and Appendix \ref{App:appendix_NEDvel}.  A slope of \ho = 70 \hounits is shown in black. }
\label{fig:localho}
\end{figure*}

In  Figure \ref{fig:localho} we plot a  local Hubble diagram for very nearby galaxies, with TRGB (left panel) and (\shoes) Cepheid (right panel) distances to \sne host galaxies as a function of velocity. The galaxies for which both Cepheid and TRGB distances have been measured are shown as black-filled circles in both plots. For nearby galaxies (v$<<$c), the redshift provides a good approximation to the Hubble velocity, but the inhomogeneous distribution of matter in the universe results in deviations from the Hubble expansion. These induced `peculiar velocities'  can be a significant component on top of the Hubble recessional velocity at low redshift.   We have corrected the observed redshifts   on a galaxy-by-galaxy basis using the linear multi-attractor model provided by NED. This model is defined by Virgo infall, the Great Attractor, and the Shapley supercluster \citep{mould_2000}. The adopted velocities are given in Table \ref{tab:trgb_calibrators}, and more detailed notes on individual galaxies and group membership, if applicable, can be found in Appendix \ref{App:appendix_NEDvel}. A line with slope of \ho = 70  \hounits and intercept of zero is shown for reference.

For the overlap sample of ten galaxies that have both TRGB and Cepheid distances measured, we note that the ratio of the dispersion about independent fits to the TRGB and Cepheid data is 0.7. Since the galaxy velocity is the same in each individual case, this indicates that the uncertainty in the distances for the TRGB method is about a factor of 1.4 smaller than for the Cepheids. A more detailed intercomparison and decomposition of the magnitude and origin of
the scatter in the two panels of Figure \ref{fig:localho} leads to a very similar
conclusion.  We fix the quadrature-summed dispersion $\sigma_{TC}$ shared by the two distance indicators, where $\sigma_C$ is the intrinsic Cepheid dispersion and $\sigma_T$ is the intrinsic TRGB dispersion.  Then $\sigma_{TC} = (\sigma_C^2 + \sigma_T^2)^{1/2} = $ 0.17 mag. Incrementally adding a peculiar motion component $\sigma_{pec}$ simultaneously to each of the two velocity-distance plots we find a best-fit solution to the data in the two diagrams with 
$\sigma_{pec} = \pm$ 130~km/sec, $\sigma_C = \pm$ 0.14~mag and $\sigma_T = \pm$0.10~mag, again confirming the higher precision of the TRGB distances in comparison to the Cepheid-based 
distances.

We note here that the Cepheid-based calibration of \ho involves greater complexity in its application relative to that of the TRGB. For instance, Cepheids possess a range of periods; exhibit a width to the Leavitt law, which is a function of the intrinsic temperature of the star; they suffer larger total and differential reddening; and the effect of metallicity on the luminosities and colors of Cepheids as a function of wavelength still remains under discussion. Currently, Cepheids beyond the LMC are too faint to have their metallicities measured directly,  requiring the use of a proxy metallicity indicator derived from young HII regions. Some studies have indicated  that there may be a break in the slope of the Leavitt law at a period of 10 days  \citep[e.g., see discussions in][]{ngeow_kanbur_2005, ngeow_2005, riess_2016}.  Possibly the most significant challenge for Cepheid measurements beyond 20 Mpc is crowding and blending from redder (RGB and AGB) disk stars, particularly for near-infrared $H$-band measurements of Cepheids. While a global fit can be applied to the entire Cepheid sample when marginalizing over these and other nuisance parameters, we note that  these issues do not apply to the single-step measurement technique used for TRGB distances.

\section{Type Ia Supernovae}
\label{sec:supernovae}

Although the details of what gives rise to \sn explosions are not well-understood from theory, the observed properties of \sne have, for many decades, been consistent with a general model in which an increase to the mass of a carbon-oxygen white dwarf in a binary system results in it approaching its Chandrasekhar mass (1.4 M$_{\odot}$), thereby triggering a thermonuclear explosion \citep{hoyle_fowler_1960, wheeler_hansen_1971}. These events are predicted to  occur if the white dwarf accretes material from a nearby companion star, or, alternatively, if the companion itself is also a white dwarf, and the merger of the two white dwarfs then triggers the explosion. Current observations seem to suggest that both  single and double degenerate types of events may occur  \citep[e.g., see the recent review by][]{livio_mazzali_2018}.

\sne have many observational advantages for determining the far-field value of \ho (as well as other cosmological parameters), and as a result  they continue to play a pivotal role in cosmology.  Primarily due to their high intrinsic luminosities, they can be observed well into the Hubble flow, where peculiar motions become a negligible contribution to their observed redshifts. In addition, because \sne are transients and fade with time, image differencing (once the \sn has faded) permits  very clean estimates of the flux, minimally affected by crowding issues. Brighter \sne have wider light curves (longer decay times) than their fainter counterparts; correction for this simple empirical fact allows these objects to be used as standardizable candles \citep{phillips_1993, riess_1998, perlmutter_1999,burns_2018}. As noted previously, for nearby, well-observed \sne  their measured dispersion amounts to only  $\pm$0.10 mag or 5\% in distance \citep{burns_2018}, depending on the filter/color combination employed. To date, no other method can provide {\it relative} distances to this precision in the redshift range 0.03 $<$ z $<$ 0.4. However, \sne are sufficiently rare and distant that determination of {\it absolute} distances requires an intermediate calibration step.

In what follows, we apply our new TRGB distances to the calibration of two published sets of \sne photometry: the Carnegie Supernova Project I sample \citep{krisciunas_2017} and the Supercal sample \citep{scolnic_2015}. We also compare the TRGB calibration to that based on the Cepheid data from \citet{riess_2016}, as recalibrated by \citet{riess_2019}. First, however, we turn our attention to potential systematic uncertainties affecting the peak luminosities of  \sne.

\subsection{Systematic Uncertainties}
\label{sec:sne_systematics}

Uncertainty remains as to whether there may be systematic effects in different \sn samples arising from  different properties of the \sn progenitors located in different environments. For instance, an empirically well-established correlation has been measured (by many different groups) between the standardized peak luminosity of the \sn (after correcting for light-curve shape and color) and the total stellar mass of the host galaxy \citep{kelly_2010, sullivan_2010, lampeitl_2010, childress_2013,uddin_2017}. The sense of the correlation is that, on average, \sne residing in more massive galaxies are more luminous (although the mass of the host galaxy is unlikely to be the actual driver of the \sne luminosity difference). Furthermore, other groups have found correlations between the standardized peak luminosity and host-galaxy specific star formation rates, age, metallicity and morphology \citep[e.g., see][and references therein]{rigault_2015}, many of which are covariant with host-galaxy mass.  

This broad issue of environment is of importance in the context of the measurement of an accurate value of \ho: if, for instance,  the properties of the local calibrating sample of galaxies differ from those of the distant sample, it could potentially introduce a systematic error in \ho. Cepheid variables are young objects found only in star-forming spiral galaxies, and cannot calibrate \sne found in elliptical or S$\emptyset$ galaxies. The distant samples of \sne, however, are found in galaxies with a much wider range of morphologies, from late- and early-type spirals, gas-poor lenticular  and elliptical galaxies having a range of masses, ages, star formation rates, and metallicities. An advantage of TRGB stars is that they are found in galaxies of all morphological types. We undertake our analysis of the TRGB calibration for the \cspi, both with and without a correction for host galaxy mass, analogous to the Cepheid calibration carried out by \citet{burns_2018}.

\subsection{The Carnegie Supernova Project I}
\label{sec:cspi}

Our primary sample of \sne for this study comes from the Carnegie Supernova Project I
\citep[\cspi;][]{krisciunas_2017}.\footnote{There have been three releases of the \cspi data: 1) \citet{contreras_2010} published data for 35 \sne; 2) \citet{stritzinger_2011} published data for 50 \sne, and 3) \citet{krisciunas_2017} have provided data for the entire \cspi set of 123
\sne. The \cspi data are available at http://csp.obs.carnegiescience.edu/data.}
The \cspi observations were obtained at Las Campanas Observatory over a five-year period, 2004-2009. The light curves for these \sne are well-sampled at $uBVgriYJH$, with most of the objects observed well before maximum. These
data form the most homogeneous and densely-sampled, multi-wavelength set of photometry measured for \sne to date. 
The high quality of the photometry \citep{krisciunas_2017} and the consistency of instruments/calibration/procedures used in the data reduction minimizes systematic differences that can be a challenge in combining multiple data sets, and makes this an ideal sample for cosmological studies. An extensive set of optical spectra were
also obtained as part of the \cspi \citep{folatelli_2013}.  The sample covers a range in redshift from  0.004 to  0.083
(1,200 to 25,000 km/sec).  The data presented in \citeauthor{krisciunas_2017}
provides a synthesis of the previous publications in the \cspi series, and
supersedes those studies.

Details of the light-curve fitting for the \cspi data using the SNooPy package
can be found in \citet{burns_2018} and references therein.  SNooPy utilizes a
color-stretch parameter (s$_{BV}$) that proves to be very effective for also
fitting fast-declining \sne, which have often been excluded in other studies. 
This is advantageous for our determination of \ho since two of our calibrators
(SN2007on and SN2011iv, in NGC~1404) are transitional between the fast decliners and normal \sne. In Table \ref{tab:sncalibphot}, we provide literature references to the optical and near-infrared (NIR) \sn photometry for the TRGB- and Cepheid-calibrator galaxies. Fits to the \cspi data, in machine-readable form, have been published by \citet{burns_2018}.

In this paper we analyze a subset of 104 (out of the total sample of 123) well-observed, high signal-to-noise objects (with light curve parameters described in \S\ref{sec:trgbcal}), five of which have TRGB distances. Of the 18 calibrating \sne used in this study, 13 have non-\cspi photometry: these have also been analyzed in a consistent manner using SNooPy. The five that have \cspi photometry are:  SN2007af, SN2007on, SN2007sr, SN2011iv, and SN2012fr. The remaining 99 (out of the sample of 104) \sne form the basis of our distant \sne sample. As discussed further in
\S\ref{sec:Ho},  calibration of these data and a determination of \ho based on
the Cepheid distances to nearby \sne hosts from \citet{riess_2016} is given in
\citet{burns_2018}.  Here we independently determine \ho using the \cspi \sne and  our new \cchp
TRGB distances.

\begin{deluxetable*}{llll}[b!]
\tablecaption{Supernovae, Hosts and References 
\label{tab:sncalibphot}} \tablecolumns{4}\tablewidth{0pt}
\tablehead{
\colhead{Name} & \colhead{Host} & \colhead{Optical Reference} &
\colhead{NIR Reference}
}
\startdata
SN1980N & NGC~1365 & \citet{hamuy_1991} & \citet{elias_1981}
\\SN1981B & NGC~4536 & \citet{tsvetkov_1982} & \citet{elias_1981}
\\SN1981D & NGC~1316 & \citet{walker_1982} & \citet{elias_1981}
\\SN1989B & NGC~3627 & \citet{wells_1994} & \citet{wells_1994}
\\SN1994D & NGC~4526 & \citet{richmond_1995} & \citet{richmond_1995}
\\SN1994ae & NGC~3370 & \citet{riess_2005} & \hspace{1cm}$\ldots$
\\SN1995al & NGC~3021 & \citet{riess_1999} & \hspace{1cm}$\ldots$
\\SN1998bu & NGC~3368 & \citet{jha_1999} & \citet{jha_1999} 
\\SN2001el & NGC~1448 & \citet{krisciunas_2003} & \citet{krisciunas_2003} 
\\SN2002fk & NGC~1309 & \citet{silverman_2012} & \citet{cartier_2014}
\\SN2006dd & NGC~1316 & \citet{stritzinger_2010} & \citet{stritzinger_2010} 
\\SN2007af & NGC~5584 & \citet{stritzinger_2011} & \citet{stritzinger_2011} 
\\SN2007on & NGC~1404 & \citet{gall_2018} & \citet{gall_2018} 
\\SN2007sr & NGC~4038 & \citet{schweizer_2008} & \citet{schweizer_2008} 
\\SN2011fe & NGC~5457 & \citet{richmond_2012} & \citet{matheson_2012} 
\\SN2011iv & NGC~1404 & \citet{gall_2018} & \citet{gall_2018} 
\\SN2012cg & NGC~4424 & \citet{marion_2016}  & \citet{marion_2016} 
\\SN2012fr & NGC~1365 & \citet{contreras_2010} &  \citet{contreras_2010}  \\
\enddata
                                        
\end{deluxetable*}

\sne are not perfect standard candles; they require three empirical corrections. The first is a stretch correction, commonly referred to as the Phillips relation \citep{phillips_1993}, which accounts for variations in the amount of radioactive $^{56}$Ni synthesized in the explosion and  which is thought to power the light-curve of the SN~Ia. Brighter events have more $^{56}$Ni, but also longer diffusion times in their ejecta, producing a correlation between peak luminosity and the width of the light-curve. The second correction uses the less well-understood  empirical correlation between the stretch-corrected peak luminosity and galaxy host mass (or metallicity) discussed in \S\ref{sec:sne_systematics}. Finally, a correction must be made to account for the extinction due to dust along the line of sight to the \sne.  

There are two commonly-used ways to deal with extinction for \sne. The first is to assume that the slope of the reddening law $R_V$ \citep{cardelli_1989, fitzpatrick_1999} is universal and work with
reddening-free magnitudes \citep{madore_1982} as originally done by \citet{tripp_1998}. In this case the reddening correction is simply a constant, $\beta$,  multiplied by the observed color of the \sne. The corrected magnitude can then be computed as:
\begin{equation}
\label{eq:m_Tripp}
B^\prime  =  B - P^1(s_{BV}-1) - P^{2}(s_{BV}-1)^2 - \beta(B-V) - 
       \alpha_M(\log_{10}M_*/M_\odot-M_0),
\end{equation}
where $P^1$ is the linear coefficient and $P^2$ is quadratic coefficient in ($s_{BV}-1$), which encapsulate the Phillips relation; $\beta$ is the slope of color correction; $B$ and $V$ are the apparent, K-corrected peak
magnitudes; and $\alpha_M$ is the slope of the correlation between peak luminosity
and host stellar mass $M_*$\footnote{ Host stellar masses are derived as 
described in \citet{burns_2018}, Appendix B.}.
This approach has the advantage of being simple and direct, but does not capture the observed diversity in $R_V$ seen in the Milky Way \citep[e.g.][]{fitzpatrick_1999} and host galaxies of \sne \citep{mandel_2011,burns_2014,nataf_2015}.
Furthermore, any intrinsic correlation in \sn color and luminosity could bias
the correction \citep{mandel_2017}. Alternatively, we can use the methods of \citet{burns_2014} to solve explicitly for the dust properties ($E(B-V)$ and $R_B$) of each 
\sn host. In this case, the corrected magnitude
becomes:
\begin{equation}
\label{eq:m_ebv}
B^\prime  =  B - P^1(s_{BV}-1) - P^{2}(s_{BV}-1)^2 - R_B E(B-V) -
       \alpha_M\left(\log_{10}M_*/M_\odot-M_0\right),
\end{equation}
where $R_B$ and $E(B-V)$ are derived from the optical and near-infrared colors of each \sn. The disadvantage of this method is that an extra step -- determining the intrinsic
colors of \sne as a function of ($s_{BV}$-1) -- must be done and errors in this
step introduce correlated errors in the values of $B^\prime$, which must be
taken into account in determining \ho.
Having made the color-dependent corrections, the apparent magnitudes of the \sne now depend only on their relative distances.

In both cases, the apparent magnitudes at maximum are computed by fitting the light-curves with SNooPy \citep{burns_2011}, which yields the time of maximum, the light-curve shape $s_{BV}$, and the magnitude at maximum for each filter. These are then used as 
inputs to a Markov Chain Monte Carlo (MCMC) fitter that simultaneously solves for all the correction factors: $P^1$, $P^2$, $\alpha_M$, $\beta$, $E(B-V)$ and
$R_B$  \citep[for full details, see][]{burns_2018}. Not only does this provide us with the corrected magnitudes, it also gives a full covariance matrix,
which is used when determining \ho and its error.

\pagebreak
\subsection{Supercal Sample}
\label{sec:supercal}

In comparison to the \cspi \sn sample, we also apply our TRGB calibration to the Supercal sample\footnote{http://kicp.uchicago.edu/~dscolnic/supercal/supercal.fitres} \citep{scolnic_2015}. This catalog merges data from five different supernova samples  (including early \cspi releases) and references them to a common photometric zero point, measured relative to the large and uniformly-calibrated area of the Pan-STARRS survey. 

Our Supercal sub-sample contains  214 objects with 0.0023 $<$ z $<$ 0.15. 
For a direct comparison with previous results, following \citet{betoule_2014}, and \citet{riess_2016}
we impose a number of cuts on the Supercal sample  by including only \sne light curves for which the SALT color parameter (c) is within $\pm$0.3, restricting the light-curve parameter (x) to be within $\pm$3.0, requiring the $\chi^2$ of the light-curve fit to be good (fitprob $>$ 0.01),  where the peak time of the light curve is constrained to better than 2 days, and where the uncertainty in the corrected peak magnitude is
$<$ 0.2 mag. As described in \citet{scolnic_2015, scolnic_2018} the \sne redshifts were corrected using a flow model, based on the observed nearby matter density.\footnote{http://cosmicflows.iap.fr/} In addition, they find and include a residual peculiar velocity error of  $\pm$250 km s$^{-1}$. For the 214 \sne in our cut sample, we find an intercept in the magnitude-redshift diagram of a$_B$ = 0.71639 (for comparison \citet{riess_2016} found an intercept of 0.71273 for a sample of 217 \sne).

\subsection{Absolute Magnitude Distribution of the Calibration \sne}
\label{sec:absmag}

To recap briefly, our TRGB sample consists of 18 calibrating \sne, and the Cepheid sample consists of 19 calibrating \sne. Ten of these \sne have both TRGB and Cepheid distances. 

The accuracy of the \ho measurement rests on the accuracy with which the average {\it absolute} magnitude, M$_{\lambda,i}$, can be determined: 
\begin{equation}
\label{eqn:absmag}
\rm M_{\lambda,i} = m_{\lambda,i} - \mu_{0,i}^{\rm TRGB/Ceph}
\end{equation}

\noindent
where $\lambda$ is the filter of the observation, i is the subsample of \sne for that filter, $m_{\lambda,i}$ is the apparent magnitude of the peak of the \sn light curve for a given filter, and  $\mu_0$ is the true calibrator distance modulus (TRGB or Cepheid). 

\begin{figure*} 
 \centering
\includegraphics[width=1.0\textwidth]{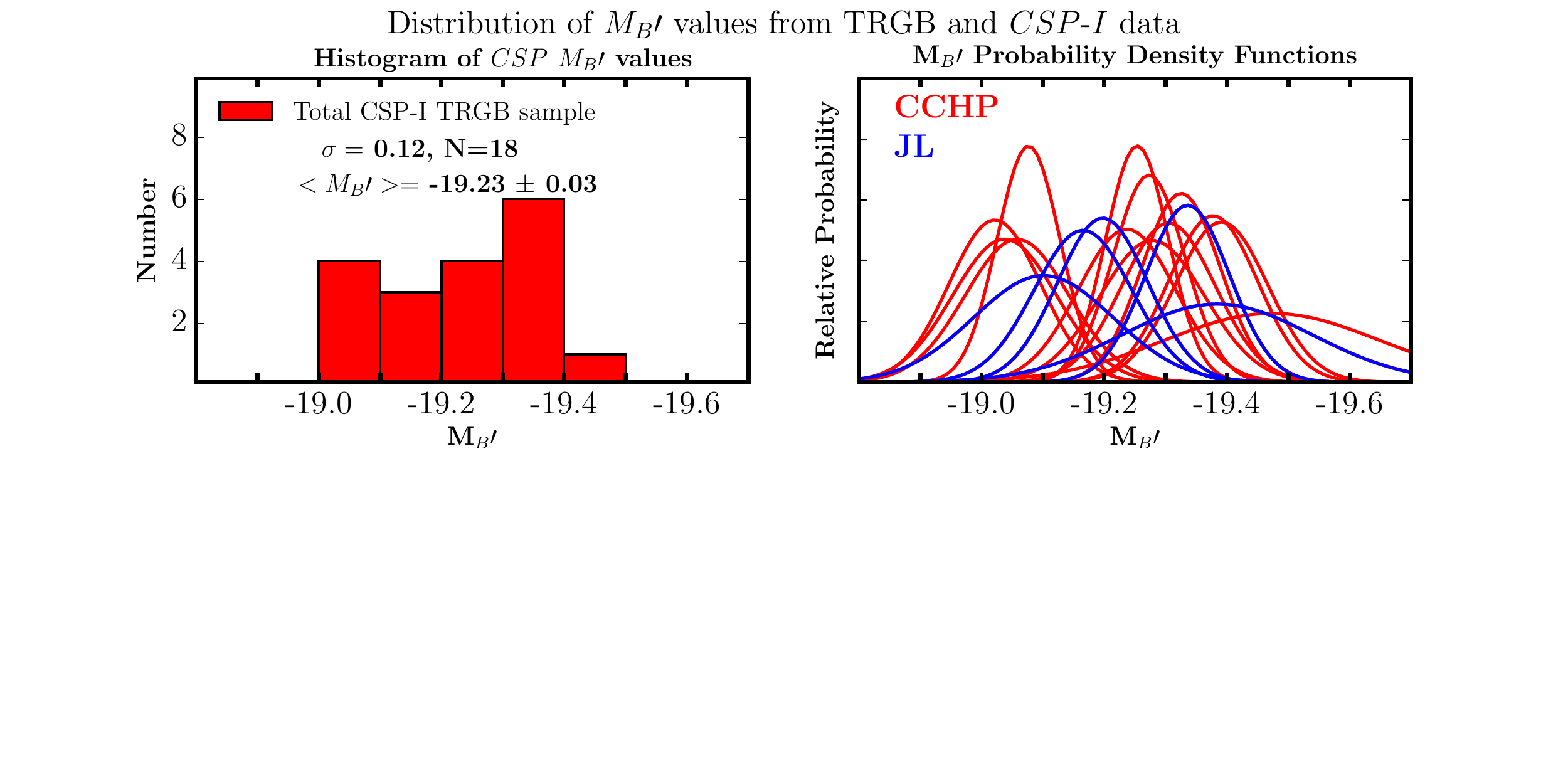}
 \caption{Tripp $B^{\prime}$-band absolute magnitude distributions for TRGB calibrators for the \cspi sample in histogram form (left-hand panel) and probability density functions (right-hand panel).   Shown in the left-hand plot are the weighted mean values for M$_{B^{\prime}}$, the dispersion, the number of galaxies, and the error on the mean. In the right-hand panel, the galaxies analyzed by \citet{jang_lee_2017b} are shown in blue. 
}
 \label{fig:MBCSP}
\end{figure*}

In Figure \ref{fig:MBCSP} we show the distribution of M$_B^\prime$ magnitudes of the TRGB calibrators for the \cspi (SNooPY-analyzed) sample of 18 \sne, in both histogram form (left-hand panel) as well as  probability density functions (right-hand panel). In the right-hand panel, the galaxies analyzed in this paper are shown in red, and those  by \citet{jang_lee_2017b} are shown in blue.  The \cspi apparent peak $B^{\prime}$  magnitudes are defined in Equation \ref{eq:m_Tripp} and listed in Column 6 of Table \ref{tab:trgb_calibrators}; the TRGB distance moduli are given in Column 3. The average weighted mean magnitude for this sample is M$_{B^{\prime}}$[N=18] = -19.225 mag with a dispersion of  $\pm$0.119 mag, and an error on the mean of $\pm$0.029 mag. The individual uncertainties in M$_{B'}$ are given by $\sigma = \sqrt{\sigma_{SN}^2 +\sigma_{TRGB}^2}$.  The relative weight (calculated as 1/$\sigma^2$) for an individual supernova is determined both by the uncertainty in its supernova photometry, $\sigma_{SN}$, as well as that in the TRGB distance, $\sigma_{TRGB}$; these uncertainties can be seen in the right-hand panel. The three objects with the lowest weights for this sample are SN1981D in NGC~1316, SN2007sr in NGC~4038, and SN2007af in NGC~5584. A series of jackknife tests show that the mean is very stable, and the difference between the weighted and unweighted means is only 0.013 mag (0.6\%). The results are very similar whether or not the two reddest objects (the Leo Group galaxies, NGC~3627 and NGC~3368)  are excluded from the sample (the weighted mean changes by only 0.004 mag or 0.2\%), and removing the transitional objects, SN 2007on and SN 2011iv, in NGC~1404 does not change the weighted mean. For the ten \sne for which there are both Cepheid and TRGB distances,  M$_{B^{\prime}}$[N=10]  = -19.239 mag with a dispersion of  $\pm$0.093 mag, and an error on the mean of $\pm$0.031 mag, amounting to a difference of 0.6\% from that for the total sample of 18 galaxies.

We next look at the \shoes sample of calibrators. We note that the \cspi $B^{\prime}$-band magnitudes (defined in Equation \ref{eq:m_Tripp} {\it cannot be compared directly} with those from \citet{riess_2016, riess_2019} since they are defined differently.\footnote{There are 19 \sne that have both $m^{CSP}_{B^{\prime}}$ and $m_B^{SC}$ magnitudes (see Table \ref{tab:trgb_calibrators}). The mean difference is 0.061 mag with a  standard deviation of 0.102 mag.} Rather, the discussion below is intended to be illustrative of where some of the differences arise in the ultimate measurement of \ho. These differences arise from  a number of factors, including differences in calibration, \sne calibrator samples, and uncertainties.

\begin{figure*} 
 \centering
\includegraphics[width=1.0\textwidth]{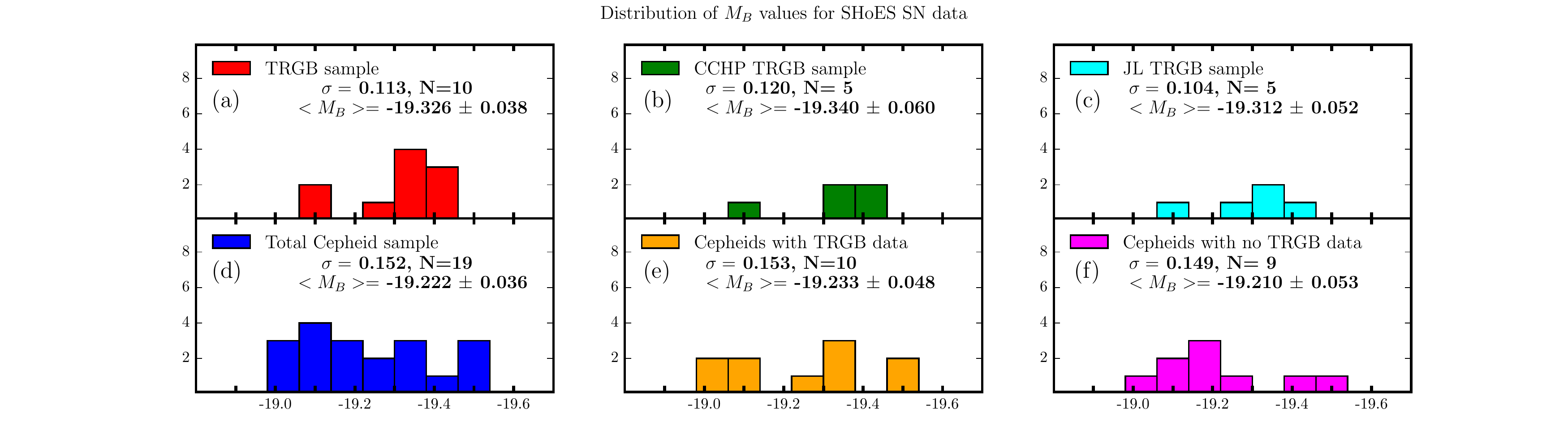}
 \caption{$B$-band absolute magnitude distributions based on the \shoes \sn data from \citet{riess_2016}, and updated to the recalibration by  \citet{riess_2019}, for different sub-samples of the calibrating galaxy samples. Figures \ref{fig:MBhist}(a-c) show the M$_{B}$ histograms based on {\it TRGB} distances for the (a) ten  galaxies having both  Cepheid distances and TRGB distances, (b) the subset of five \cchp TRGB distances, and (c) the remaining subset of five TRGB galaxies measured by \citet{jang_lee_2017b}.  Figures \ref{fig:MBhist}(d-f) show the histograms based on the  \citet{riess_2016, riess_2019} {\it Cepheid}  distances for (d) their full sample of sample of 19  galaxies, (e) the overlap sample of ten  galaxies having measured Cepheid distances and TRGB distances, and (f) the non-overlap sample of galaxies with Cepheid distances, but no TRGB distances.   Shown in the plots are the weighted mean values for M$_B$, the dispersion, the number of galaxies, and the error on the mean. We note that the dispersion in the TRGB calibration of the \sne is significantly smaller ($\sigma = \pm$0.10-0.12) than for the Cepheid calibrators ($\sigma = \pm$0.15), and is the same as the far-field dispersion of 0.1 mag.
}
 \label{fig:MBhist}
\end{figure*}

In Figure \ref{fig:MBhist} we show the \sne M$_B$ absolute magnitude distributions (M$_{B,i}$ = m$_{b,i} - \mu_{0,i}^{\rm TRGB/Ceph}$) based on the \shoes \sn data for various subsets of the TRGB and Cepheid calibrating galaxy samples. In this figure, the absolute magnitudes were calculated based on the apparent \sne m$_b^{max}$ magnitudes listed in Column 6 of Table \ref{tab:trgb_calibrators} (from \citet{riess_2016}, Table 5); the Cepheid distance moduli  listed in Column 8 of Table \ref{tab:trgb_calibrators} (also from \citet{riess_2016}, Table 5), and the TRGB distance moduli from the present paper listed in Column 3 of Table \ref{tab:trgb_calibrators}.   We update these distance moduli from the  \citet{riess_2016} to the \citet{riess_2019} magnitude scale.  The \citet{riess_2016} distances and \ho values are tied to three anchors: the Milky Way, NGC~4258 and the LMC, which resulted in a value of \ho = 73.24 \hounits. For the purpose of comparison in calculating  M$_{B,i}$ values, we compare the TRGB and Cepheid distances and \ho values anchored to the LMC alone. The \citet{riess_2019} \ho value for the LMC alone is 74.22 \hounits, a difference of 5log$_{10}(73.24/74.22)$ =  -0.029 mag. In Figures \ref{fig:MBhist}(d-f) this difference is added to the Cepheid moduli from Table \ref{tab:trgb_calibrators}. 

Figures \ref{fig:MBhist}(a-c) show the M$_{B}$ histograms based on {\it TRGB} distances for the (a) ten  galaxies having measured Cepheid distances and TRGB distances, (b) the subset of five \cchp TRGB distances, and (c) the remaining subset of five TRGB galaxies measured by \citet{jang_lee_2017b}. The weighted mean absolute  M$_{B}$ magnitudes are all consistent to within the quoted errors; these values are displayed in each of the panels, along with the standard deviations, $\sigma$, and numbers of objects, N. Figures \ref{fig:MBhist}(d-f) show the histograms based on the  \citet{riess_2016} {\it Cepheid}  distances for (d) their full sample of sample of 19  galaxies, (e) the overlap sample of ten  galaxies having measured Cepheid distances and TRGB distances, and (f) the non-overlap sample of galaxies with Cepheid distances, but no TRGB distances. The weighted mean absolute  M$_{B}$ magnitudes are again all internally consistent to within the quoted errors, labeled in each of the panels. However, for the sample of 10 overlapping galaxies (panels (a) and (e)), the TRGB (-19.326 $\pm$ 0.038 mag) and Cepheid calibrations (-19.233 $\pm$ 0.048 mag) of M$_{B}$ differ by 0.104 mag (4.4\% in distance). These values diverge  as a result of the combination of: (1)  the difference in zero-point calibration, (2) differences in the relative weightings of TRGB and Cepheid distances, (3)  differences between the full sample of 19 galaxies compared to the overlap sample of 10 galaxies and (4) differences in the \sne apparent $B$ magnitudes.

\section{Determination of the Hubble Constant}
\label{sec:Ho}

We now apply our TRGB calibration to the \cspi \sne sample, and determine a value for \ho using an MCMC method. We test the robustness of this value by examining different subsets of the calibrators and distant \sn data sets. We list the values of \ho for these different samples in Table \ref{tab:snoopyhoerrors}. In \S\ref{sec:shoes} we then apply our TRGB calibration to the Supercal \sne sample. We compare the TRGB and Cepheid calibrations for both the \cspi and Supercal far-field \sn samples. Finally, we return to a  discussion of Table \ref{tab:trgbhoerrors} in the context of our overall error budget in \S\ref{sec:ho_summary}. 

\subsection{TRGB Calibration of the CSP I SN Ia Sample}
\label{sec:trgbcal}

In this section, we use the measured TRGB distances to our sample of 18 nearby SN Ia calibrators, described above, and listed in Table \ref{tab:trgb_calibrators}, and apply this zero-point calibration to the far-field \cspi sample of \sne. References for the supernova photometry for the TRGB calibrating sample are given in Table \ref{tab:sncalibphot}; these \sne have been analyzed using SNooPy, consistent with the distant \cspi sample. We focus our analysis on the sample of \sne with s$_{BV} >$ 0.5 and $E(B-V) <$
0.5~mag, avoiding the reddest and fastest decliners, of which there are 18 (10 with s$_{BV} < $ 0.5 and
8 with $E(B-V) > 0.5$~mag).  The observed $B$ magnitudes are modeled with Equation~\ref{eq:m_Tripp} where $\mu$ = $\mu(z,H_0)$ for the distant sample, and $\mu$ = $\mu^0_{TRGB}$ for the nearby calibrators, and where \ho is treated as a free parameter. An intrinsic error term, $\sigma^2$, a peculiar velocity term, and a zero-point error for the non-\cspi photometry, have all been added in quadrature with the individual galaxy statistical errors from Table \ref{tab:trgb_calibrators}. We carried out the analysis both with and without a host-galaxy-mass correction, as described in detail in \citet{burns_2018}. Our covariance matrices for this analysis are published online. 

As in \citet{freedman_2009},  we use a simple Taylor series expansion to derive the luminosity distance and the distance modulus as a function of redshift \citep{visser_2004, caldwell_kamionkowski_2004}. This kinematic model parameterization  is independent of theoretical assumptions about the dark matter and energy content of the universe. The first- and second-order terms in this expansion are parameterized by the deceleration parameter, 
q$_0$ = ${ {-\ddot{a}a} \over{\dot{a}^2} }$ and the cosmic jerk, j$_0$ = 
${ {-\dddot{a}a^2} \over{\dot{a}^3} }$, which is the third derivative of the scale factor. The luminosity distance can then be written as:

\begin{equation}
\label{eq:dlum}
d_L(z,z_{hel},H_0,q_0,j_0) = {{(1+z_{hel})} cz \over{{(1+z)}H_0}} \lbrace 1 + {1\over{2}}[1-q_0]z - {1\over{6}} [1 - q_0 - 3q_0^2 + j_0]z^2\rbrace 
\end{equation} \linebreak

\noindent Our redshifts are sufficiently small that the 3rd-order $j_0$ term is unimportant. We assume a flat cosmology, $\Omega_k$ = 0,  with $q_0 = \Omega_m/2 - \Omega_{DE} = -0.53$, where $\Omega_m$ = 0.315 \citep{planck_2018}.
The distance modulus  is then given by:
\begin{equation}
\label{eq:mu}
\mu(z,H_0,q_0) =  {\rm 5~log_{10}} \lbrace  {{(1+z_{hel})} cz \over{{(1+z)}H_0}} (1 + {1-q_0\over{2}}z )\rbrace  + 25.
\end{equation} \linebreak
\noindent

In Table \ref{tab:snoopyhoerrors} we present a summary of the values of \ho derived from the Tripp and from the explicit reddening-corrected magnitudes, individually for each of $BiJH$ \cspi filters. We analyze separately the cases where there is a correction for host mass (HM) and no correction (noHM), for different color and stretch constraints, as labeled.  Also listed, in the final column, is the number  of  calibrating \sne  observed  in  each  filter.  The tabulated uncertainties are those determined from a diagonal covariance matrix with respect to the  TRGB distances.

\begin{deluxetable*}{l|llll|llll|l}
\tablewidth{0pc}
\tablecaption{Best-fit values of $H_0$ in 
              $\mathrm{km\cdot s^{-1}\cdot Mpc^{-1}}$ \label{tab:snoopyhoerrors}}
\tablehead{
 & \multicolumn{4}{c|}{Tripp} & 
\multicolumn{4}{c|}{$E(B-V)$} & \\
filters & $H_0$ (HM) & $\sigma$ & $H_0$ (noHM) & $\sigma$ &
          $H_0$ (HM) & $\sigma$ & $H_0$ (noHM) & $\sigma$ &
$N_{calib}$}
\startdata
\multicolumn{10}{c}{Full Sample} \\
B & $69.70$ & $1.41$ & $70.32$ & $1.32$ & $70.63$ & $1.34$ & $71.22$ & $1.29$ & 18 \\
i & $69.00$ & $1.30$ & $69.73$ & $1.23$ & $68.90$ & $1.36$ & $69.56$ & $1.32$ & 16 \\
J & $69.56$ & $1.36$ & $69.98$ & $1.35$ & $69.78$ & $1.51$ & $70.14$ & $1.48$ & 16 \\
H & $69.21$ & $1.35$ & $69.69$ & $1.30$ & $69.46$ & $1.44$ & $69.88$ & $1.40$ & 16 \\
\multicolumn{10}{c}{$s_{BV} > 0.5$} \\
B & $69.65$ & $1.37$ & $70.13$ & $1.31$ & $70.42$ & $1.18$ & $70.91$ & $1.19$ & 18 \\
i & $68.92$ & $1.27$ & $69.54$ & $1.21$ & $68.85$ & $1.32$ & $69.43$ & $1.29$ & 16 \\
J & $69.43$ & $1.34$ & $69.74$ & $1.32$ & $69.59$ & $1.49$ & $67.97$ & $1.44$ & 16 \\
H & $69.21$ & $1.33$ & $69.66$ & $1.32$ & $69.46$ & $1.43$ & $69.88$ & $1.39$ & 16 \\
\multicolumn{10}{c}{$E(B-V) < 0.5$} \\
B & $69.67$ & $1.34$ & $70.28$ & $1.28$ & $70.16$ & $1.33$ & $70.85$ & $1.27$ & 18 \\
i & $69.00$ & $1.30$ & $69.68$ & $1.22$ & $68.43$ & $1.36$ & $69.30$ & $1.32$ & 16 \\
J & $69.46$ & $1.36$ & $69.72$ & $1.35$ & $69.28$ & $1.50$ & $69.69$ & $1.48$ & 16 \\
H & $69.12$ & $1.34$ & $69.55$ & $1.32$ & $68.90$ & $1.48$ & $69.47$ & $1.41$ & 16 \\
\multicolumn{10}{c}{$E(B-V) < 0.25$} \\
B & $68.83$ & $1.42$ & $69.48$ & $1.34$ & $70.77$ & $1.43$ & $71.42$ & $1.36$ & 15 \\
i & $68.70$ & $1.41$ & $69.58$ & $1.32$ & $69.21$ & $1.48$ & $70.03$ & $1.41$ & 13 \\
J & $69.79$ & $1.48$ & $70.16$ & $1.46$ & $70.20$ & $1.64$ & $70.51$ & $1.56$ & 13 \\
H & $69.79$ & $1.49$ & $70.21$ & $1.47$ & $69.82$ & $1.60$ & $70.29$ & $1.56$ & 13 \\
\multicolumn{10}{c}{$s_{BV} > 0.5$ and $E(B-V) < 0.5$} \\
B & $69.67$ & $1.35$ & $70.27$ & $1.29$ & $69.88$ & $1.17$ & $70.46$ & $1.18$ & 18 \\
i & $68.98$ & $1.29$ & $69.66$ & $1.21$ & $68.35$ & $1.30$ & $69.09$ & $1.25$ & 16 \\
J & $69.45$ & $1.36$ & $69.70$ & $1.35$ & $69.10$ & $1.48$ & $69.46$ & $1.46$ & 16 \\
H & $69.15$ & $1.35$ & $69.57$ & $1.32$ & $68.87$ & $1.45$ & $69.39$ & $1.42$ & 16 \\
\enddata
\end{deluxetable*}

For the  $(B-V)$ color-corrected Tripp $B$-band analysis with $s_{BV} > 0.5$ and $E(B-V) < 0.5$, we find \ho = 69.67 $\pm$ 1.35 \hounits and  \ho = 70.27 $\pm$ 1.29 \hounits, with and without the mass correction, respectively.  For the explicit reddening-correction ($E(B-V)$) model, we find \ho = 69.88 $\pm$ 1.17 \hounits and  \ho = 70.46 $\pm$ 1.18 \hounits,  respectively, with all values agreeing well to within their quoted uncertainties (see Table \ref{tab:snoopyhoerrors}). 

\citet{burns_2018} found that the largest source of systematic error for the \sne was the difference in average host galaxy stellar mass between the \citet{riess_2016} Cepheid sample and the more distant \cspi sample.   In that case, the  limited  mass  range  of  the calibrating sample introduced a covariance between \ho and $\alpha_M$. It limited the precision with which $\alpha_M$ could be estimated, in turn increasing the overall \ho systematic error, unlike the case for the TRGB sample. The agreement cited in the paragraph above underscores the additional advantage of this analysis in that the galaxy host masses of the TRGB sample are more massive on average than the Cepheid sample, making the TRGB set a good match to the \cspi distant sample.

Based on this MCMC analysis, our adopted best value for \ho is  69.8 $\pm$ 1.3 \hounits. This value is obtained using the $B$-band photometry, for which the largest number of calibrators is available, and by conservatively restricting the sample to \sne that are not fast decliners ($s_{BV} > 0.5$) and not highly reddened ($E(B-V) < 0.5)$; allowing for a host-mass correction; and averaging the values from the Tripp derivation (\ho = 69.67 $\pm$ 1.35 \hounits) and the explicit  $E(B-V)$ reddening correction (\ho = 69.88 $\pm$ 1.17 \hounits). We note that most of the values within this table fall within 1\sig of our adopted value, despite  differences in sample size, wavelength, and whether or not a host-mass correction is applied. Most of the outliers occur at wavelengths where there are only a small number of calibrators. We note  that a direct reddening correction results in a decrease of the uncertainty in the $B$-band solution, while  the uncertainty increases in the case of the redder $JH$-bands.

The systematic errors in the {\it zero point} have been listed separately in Table \ref{tab:lmc_systematics}. We combine in quadrature the systematic error for the zero point with that from the \sne analysis  to give the total uncertainty in \ho. The final adopted errors are summarized in Table \ref{tab:trgbhoerrors} of \S\ref{sec:hubdiag} below. Our adopted value of the Hubble constant and its uncertainty, as derived from the TRGB method applied to the \cspi sample of \sne then becomes \ho = 69.8 $\pm$ 0.8 (stat) $\pm$ 1.8 (sys) \hounits.

\begin{deluxetable*}{lccc}
\tablecaption{Summary of \ho Uncertainties \label{tab:trgbhoerrors}} 
\tablehead{\colhead{Source of Error} & \colhead{Random Error} & \colhead{Systematic Error} & \colhead{Description}} 
\startdata
LMC Zero Point       &       1.0\%       &        1.8\%      &      \S\ref{subsec:TRGB_calib}  \\
\cspi \sne &       0.5\%       &        1.8\%      &   \S\ref{sec:trgbcal}    \\         
 \hline
Total &             1.1\%       &        2.5\%   &  In quadrature \\
\enddata
\end{deluxetable*}

\subsection{The Hubble Diagram}
\label{sec:hubdiag}

\begin{figure*} 
 \centering
\includegraphics[width=1.0\textwidth]{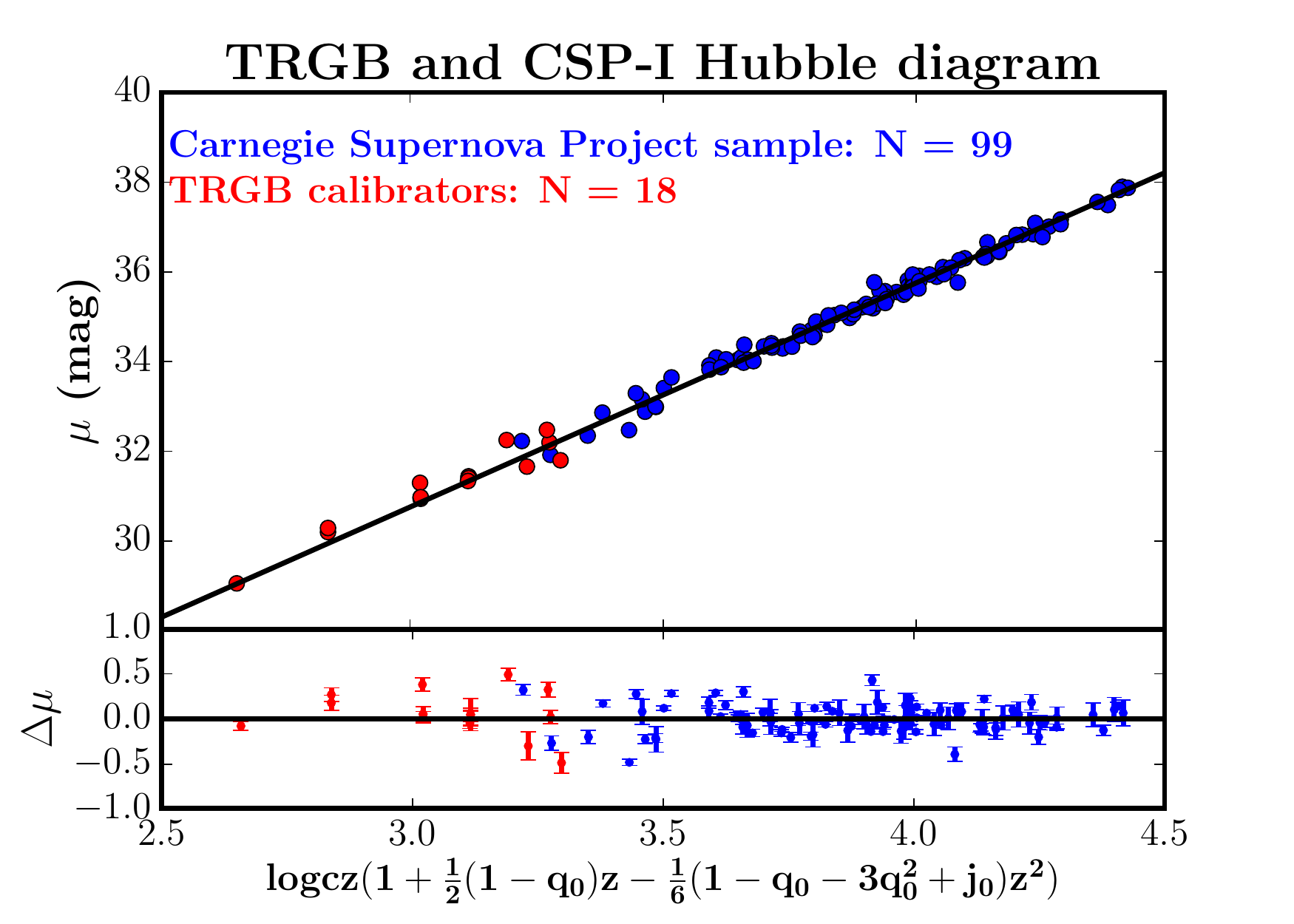}
 \caption{A Hubble diagram for 99 \sne observed as part of the CSP I (blue squares). Shown also are the TRGB calibrators (this paper, red circles). A slope = 5 line is plotted. The TRGB  velocities have been corrected as described in \S \ref{sec:nearbyHo}. }
\label{fig:CSPHubble}
\end{figure*}

In Figure \ref{fig:CSPHubble} we show the Hubble diagram for the 18 TRGB calibrating galaxies (red filled circles) at low redshift connecting to the CSP sample of 99 SNe Ia (blue filled squares) extending out to z = 0.08. The distance moduli are computed from Tripp $B$ magnitudes  for \sne with $s_{BV}>0.5$ and $E(B-V)<0.5$. A host mass correction has been applied. Residuals from the fit in distance modulus  are shown in the lower panel. This figure illustrates the major result of this paper: a new and completely independent calibration of \ho based on the TRGB and \cspi \sne.

\begin{figure*} 
 \centering
\includegraphics[width=1.0\textwidth]{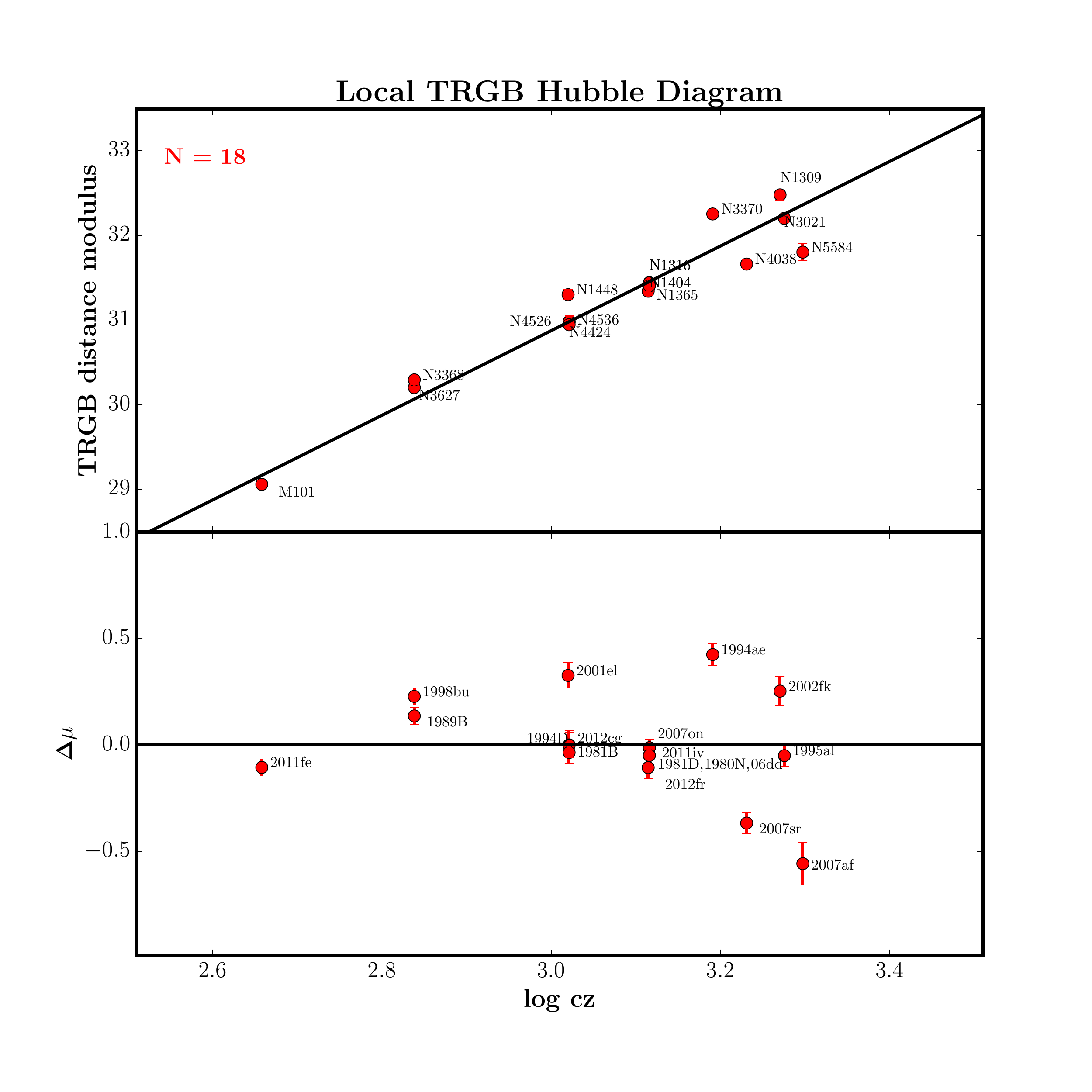}
 \caption{Upper panel. A Hubble diagram for the total sample of 18 TRGB calibrators where each point is labeled by its host-galaxy name. The slope is determined from the distant CSP sample of 99 galaxies. Lower panel. Residuals from the Hubble diagram. Each point is labeled by its supernova name. Once again, the velocities have been corrected using the NED nearby-galaxies mass model.}
\label{fig:localtrgb}
\end{figure*}

In Figure \ref{fig:localtrgb} the nearby TRGB calibrators are shown in more detail. The points are labeled by galaxy name in the Hubble diagram shown in the upper plot, and by supernova name in the lower panel showing the residuals from the Hubble diagram. A line with a slope of five is shown in the upper plot.

\begin{figure*} 
 \centering
\includegraphics[width=1.0\textwidth]{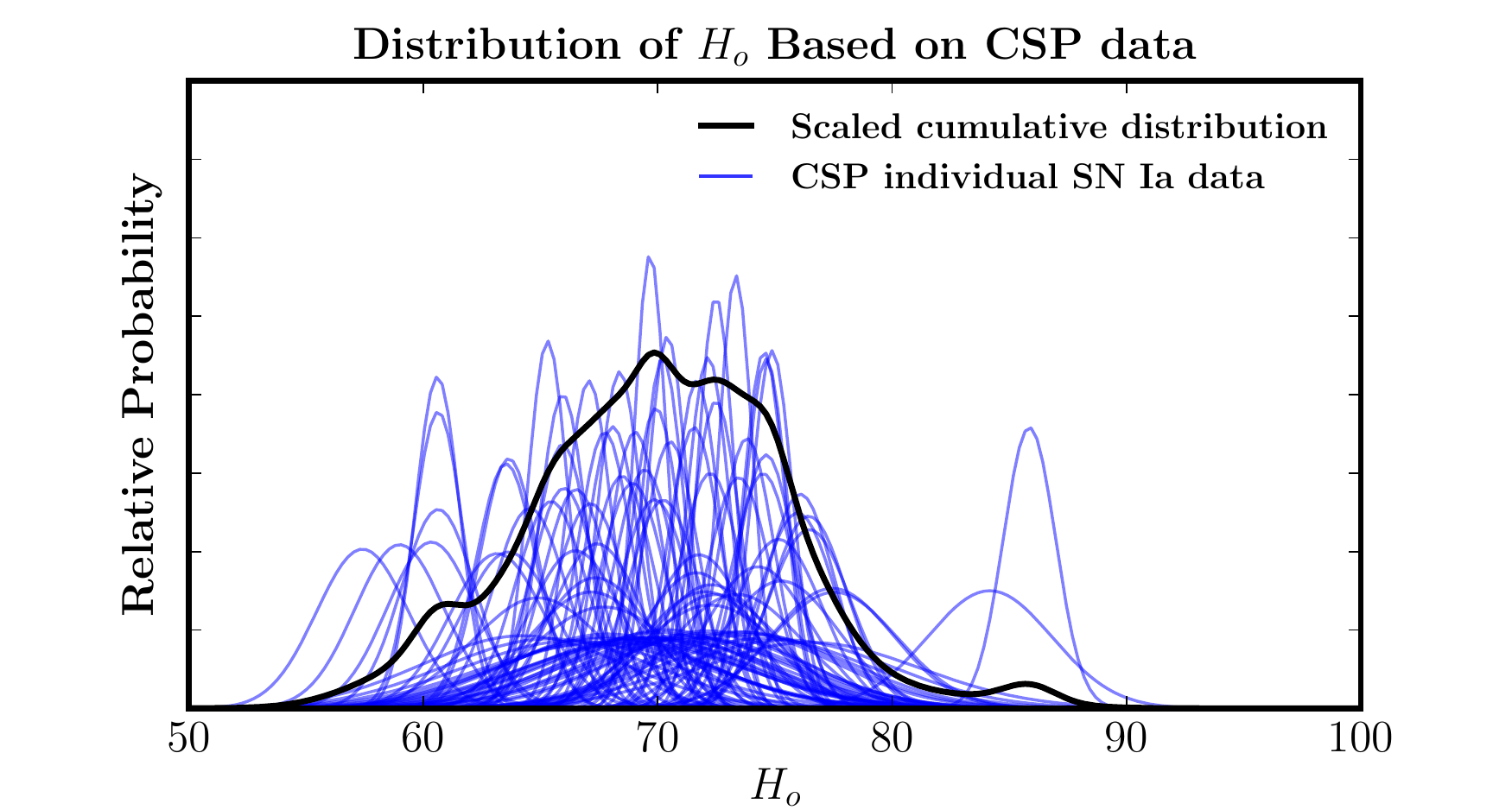}
 \caption{Probability distribution functions for the individual values of \ho calculated for each \cspi \sn are shown in blue.  The summed and scaled probability density function is shown in black. 
 }
 \label{fig:CSPpdfs}
\end{figure*}

\begin{figure*} 
 \centering
\includegraphics[width=1.0\textwidth]{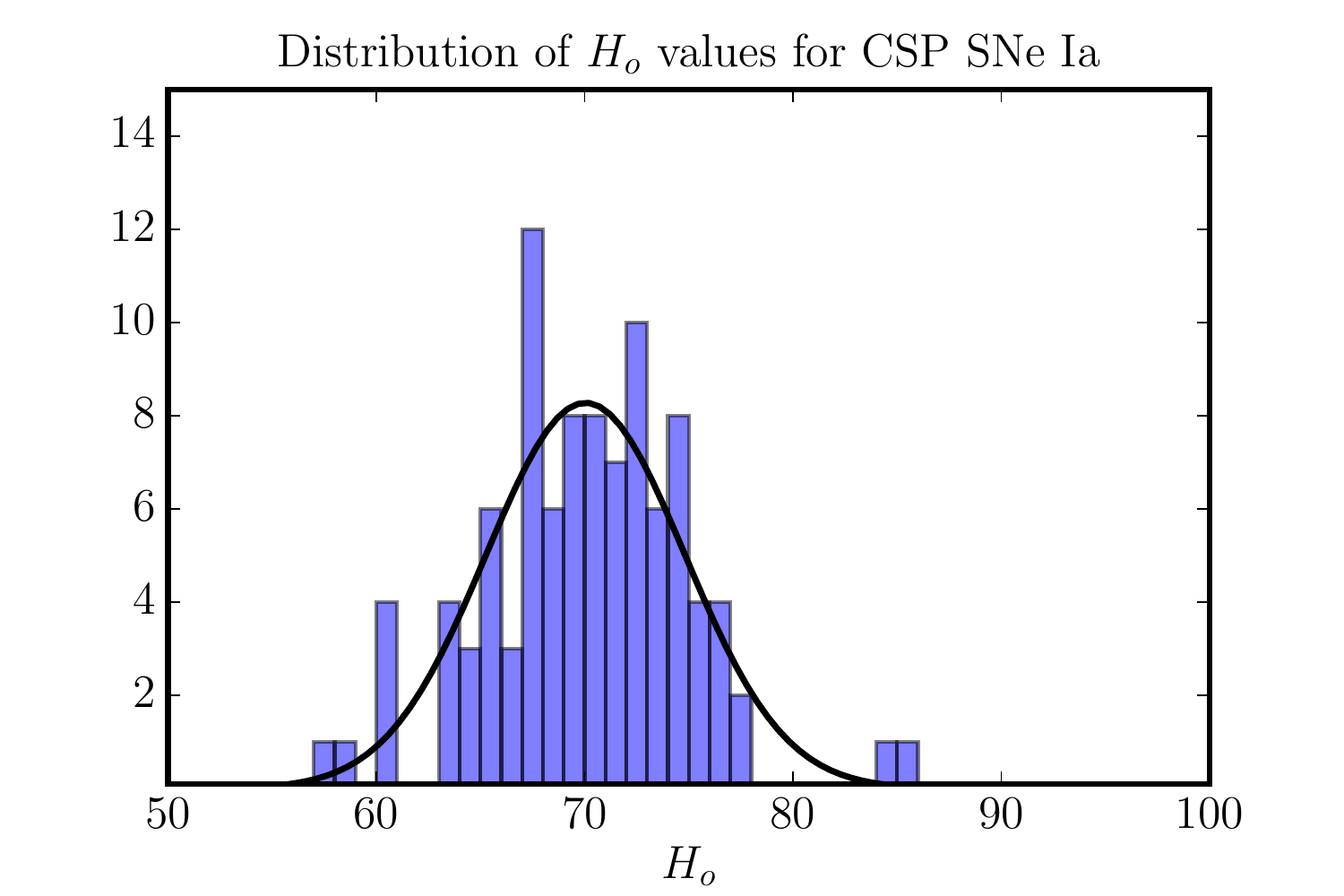}
 \caption{A histogram of \ho values for the individual 99 \sne observed as part of the \cspi, calibrated by the TRGB distances from this paper. }
 \label{fig:CSPhist}
\end{figure*}

The distribution of \ho values calculated for each of the individual \sne is shown in Figure \ref{fig:CSPpdfs}. The individual \ho values and their 1\sig uncertainties are plotted for the 99 \cspi \sne as individual Gaussians in blue. The black line is the summed and scaled probability density function. In Figure \ref{fig:CSPhist}   the same \ho values are shown in histogram form.

Our  result agrees to within 1.3\sig of the combined uncertainties with that of \citet{burns_2018} (accounting for the new \citet{pietrzynski_2019} LMC calibration), who obtained a value of \ho = 73.2 $\pm$ 2.3 \hounits. In \citet{burns_2018}, the \cspi data were calibrated with the published Cepheid distances from \citet{riess_2016}, but independently analyzed, and they resulted in a larger estimated uncertainty for \ho than obtained by \citet{riess_2016}.

\subsection{Comparison with the Supercal Sample}
\label{sec:shoes}

\begin{figure*} 
 \centering
\includegraphics[width=1.0\textwidth]{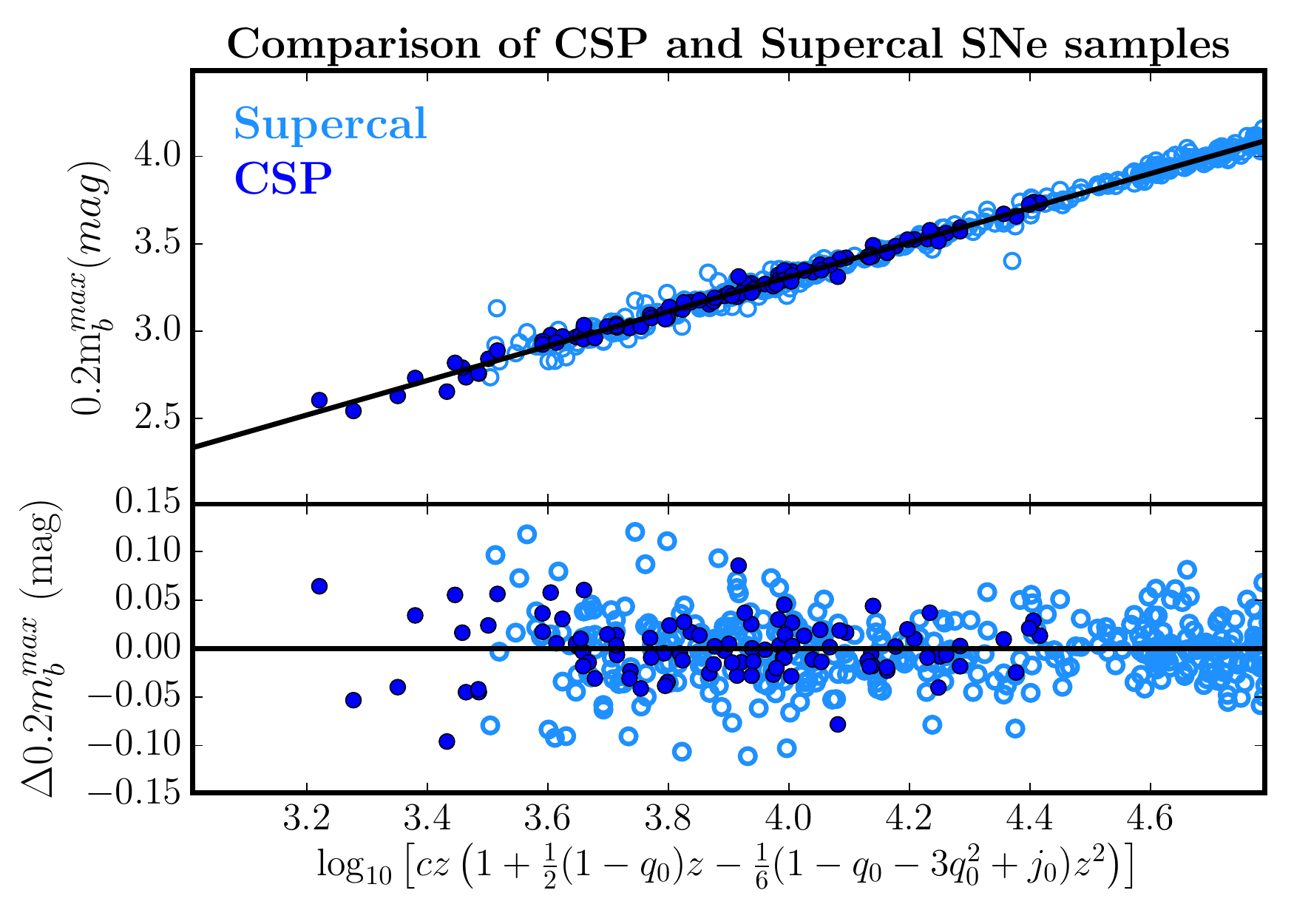}
 \caption{Upper panel. A Hubble diagram for 99 \sne observed as part of the \cspi (blue filled circles). Shown also is the Supercal sample of 214 \sne (light blue open circles). The two data sets have been arbitrarily set to the same zero point for comparison purposes only. Lower panel. Residuals from the Hubble diagram above. Colors remain the same as above. }
 \label{fig:SHOESCSPHubble}
\end{figure*}

In Figure \ref{fig:SHOESCSPHubble} we show a combined Hubble diagram including  the \cspi sample of 99 SNe Ia (blue filled circles) from \citet{burns_2018} superimposed on the Supercal sample of 214 galaxies (lighter blue open circles) from \citet{scolnic_2015} with 0.023 $<$ z $<$ 0.15, as described in \S\ref{sec:supercal}. Plotted is 0.2m$_b^{max}$ (where m$_b^{max}$ is the apparent $B$ magnitude of the \sn at maximum luminosity) versus redshift.  This comparison illustrates the high quality of the \cspi sample photometry.

Applying our TRGB zero-point calibration to the Supercal \sne sample,  we find 
a value of \ho = 70.4 $\pm$ 1.4 \hounits (not including the systematic uncertainty). This estimate is based on the  ten \sne in Table 5 of \citet{riess_2016} for which there is $B$-band \sn photometry, and for which we have TRGB distances.\footnote{For the sample of 19 Cepheid galaxies, we find \ho = 73.83 \hounits.} This value can be directly compared with the published value of \ho = 74.22 $\pm$ 1.82  \hounits from \citet{riess_2019}, using only the LMC as an anchor galaxy. In both cases the recent \citet{pietrzynski_2019} distance modulus of 18.477 mag to the LMC has been adopted. (For comparison, the older \citet{riess_2016} calibration used the \citet{pietrzynski_2013} distance modulus to the LMC of 18.493 mag, and obtained an LMC-only anchored value of \ho = 72.04 $\pm$ 2.56 \hounits.) This new LMC distance modulus contributes   0.7\% (0.016 mag) to the 3\% difference between the \citet{riess_2016} and \citet{riess_2019} LMC-anchored values of \ho.   The dominant contribution to the difference comes from the \citet{riess_2019} recalibration of the $WFC3$ photometry for the LMC Cepheids.

\begin{figure*} 
 \centering
\includegraphics[width=1.0\textwidth]{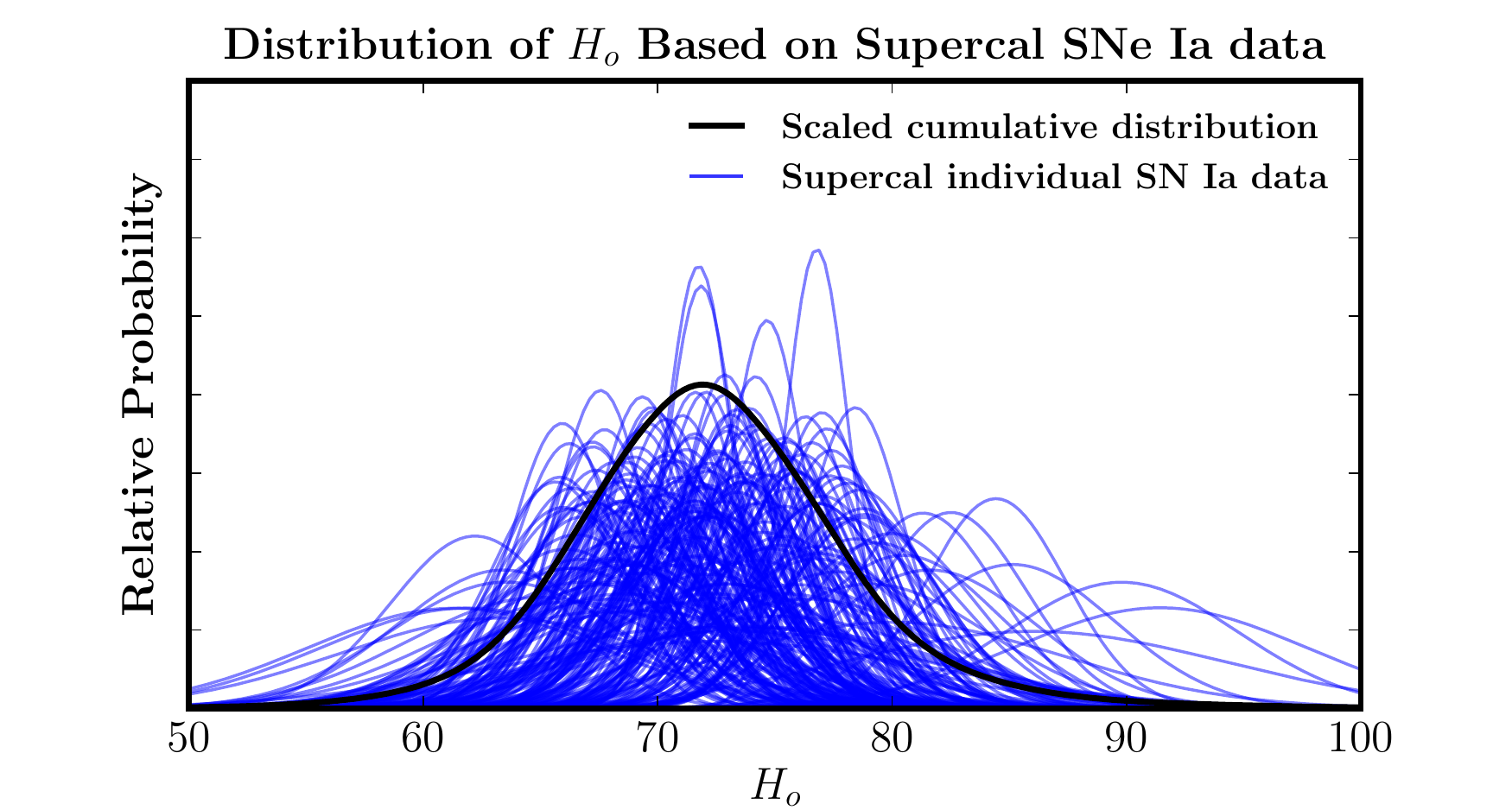}
 \caption{Probability distribution functions for the individual values of \ho calculated for each Supercal \sn are shown in blue.  The scaled and summed probability density function is shown in black.}
 \label{fig:Scolnicpdfs}
\end{figure*}

\begin{figure*} 
 \centering
\includegraphics[width=1.0\textwidth]{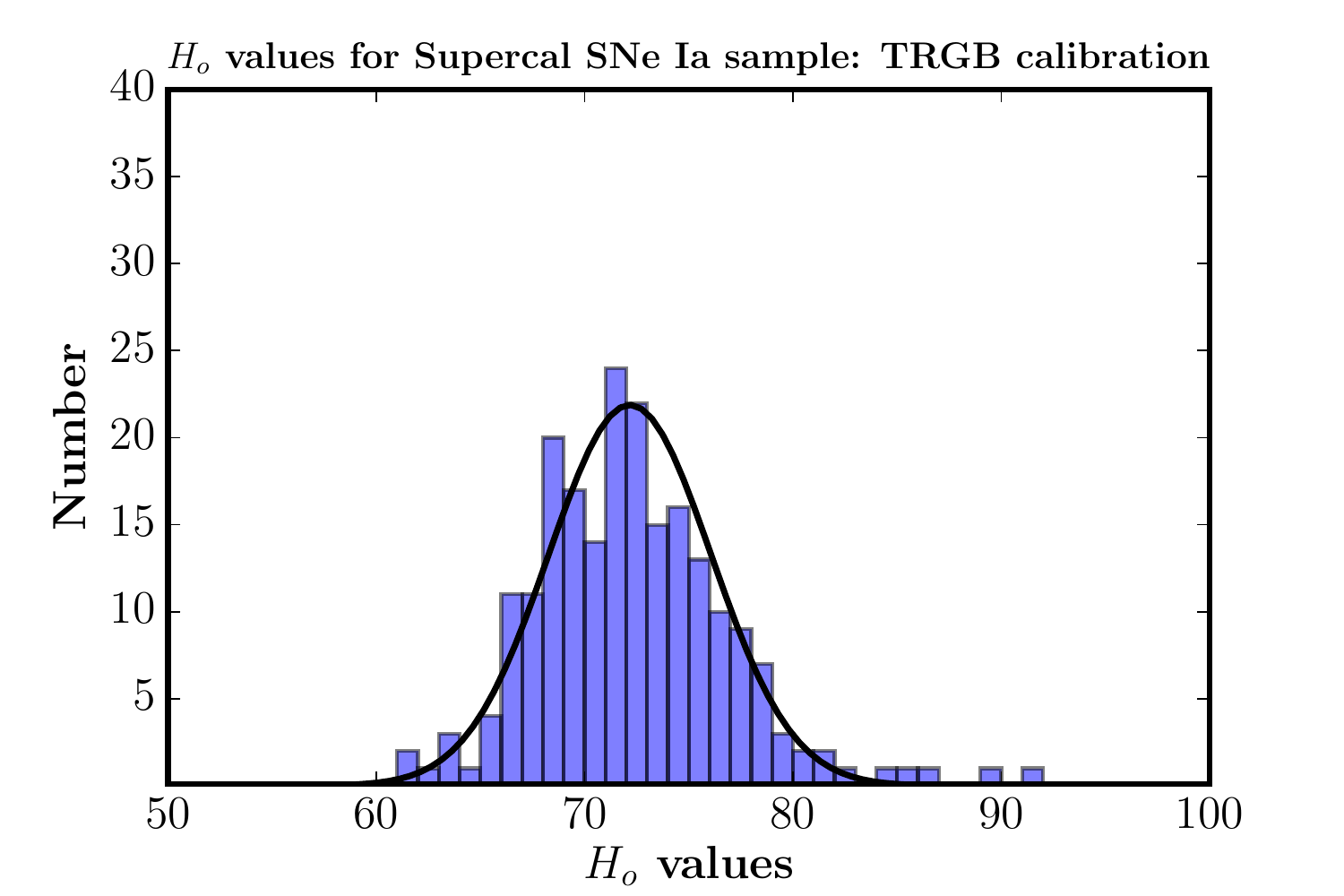}
 \caption{A histogram of \ho values for the individual 214 \sne observed as part of Supercal, calibrated by the TRGB distances from this paper.}
 \label{fig:Scolnichist}
\end{figure*}

The distribution of \ho values from the Supercal data is shown in Figure \ref{fig:Scolnicpdfs}. \ho values and 1\sig uncertainties for the 214 Supercal \sne are plotted as individual Gaussians in blue. A scaled and summed probability density function is given by the black line. In Figure \ref{fig:Scolnichist}  the \ho values are shown in histogram form.

\pagebreak
\subsection{Summary of H$_0$ Results}
\label{sec:ho_summary}

As described in \S\ref{sec:trgbcal}, our best value of the Hubble constant, based on the TRGB distances to 18 \sne anchored to 99 \cspi distant \sne, and determined via an MCMC analysis, results in a value of \ho = 69.8 $\pm$ 0.8 (stat) $\pm$ 1.7 (sys) \hounits. 

Our result differs from the best value of \ho = 74.03 $\pm$ 1.42 \hounits from \citet{riess_2019} for a number of reasons. These include the zero-point calibration, different sub-samples of galaxies that contribute to the determination of the absolute \sne magnitudes, different uncertainties for individual galaxies distances and \sn photometry (and therefore different weightings), different \sne samples (\cspi versus Supercal), different treatment of \sne host-galaxy mass, and different analysis methods (MCMC versus a maximum likelihood/matrix inversion approach). The difference (at the 1.7\sig level)  suggests that further work is necessary to reduce the remaining systematic uncertainties.

\subsection{Comparison With Other Recent Measurements of the Hubble Constant}
\label{sec:others}

Recent measurements of the local expansion rate have been made based on a number of alternative methods including strong gravitational lensing \citep{suyu_2017,birrer_2018}, the Tully-Fisher relation using the TRGB \citep{mould_sakai_2008} or Cepheids \citep{sorce_2013}, the optical counterpart to GW170817 \citep{abbott_2017}, in addition to the Cepheid calibration of \sne \citep{riess_2011,freedman_2012, riess_2016, burns_2018}. All of these studies find \ho values in the range of 70-74 \hounits, with individual  uncertainties quoted in the 3-10\% (2-7 \hounits) range. None of them meaningfully overlap the Planck result. We note also that none of them fall on the low side of the Planck result, as would be expected if they were randomly sampled measurements and the Planck \ho value were the `true' \ho value. 

A number of re-analyses of the Cepheid/\sne data from \citet{riess_2011} and \citet{riess_2016} have also been carried out 
\citep{efstathiou_2014,  zhang_2017,feeney_2018}. The \citet{zhang_2017} study differs in approach from these other studies in that it was carried out blinded. The starting point for these analyses, in all cases, is the \hst Cepheid sample reduced by \citet[][and references therein]{riess_2016}.  The \citet{burns_2018} study also uses the \citet{riess_2016} Cepheid distances for its zero-point calibration.  The recent analyses of \citet{zhang_2017,feeney_2018} and \citet{burns_2018} all utilize a Bayesian MCMC approach to estimate the uncertainties in \ho.  Although differing in their details, the conclusion of all of these studies is that the covariance amongst the multiple parameters in the MCMC analyses is actually very small -- with a single exception: that of the distances to the nearby calibrators (Milky Way, NGC\, 4258, M31 and the LMC), which set the zero point  \citep[e.g.,][]{zhang_2017,feeney_2018}. This conclusion highlights the (simple and obvious) point that can be made without any formal re-analysis: the values of \ho move in lock step with the adopted zero point of the Leavitt law.

\subsection{Comparison of H$_0$ Values for Cepheids, TRGB and Planck}
\label{sec:local_planck}

\begin{figure*} 
 \centering
\includegraphics[width=1.0\textwidth]{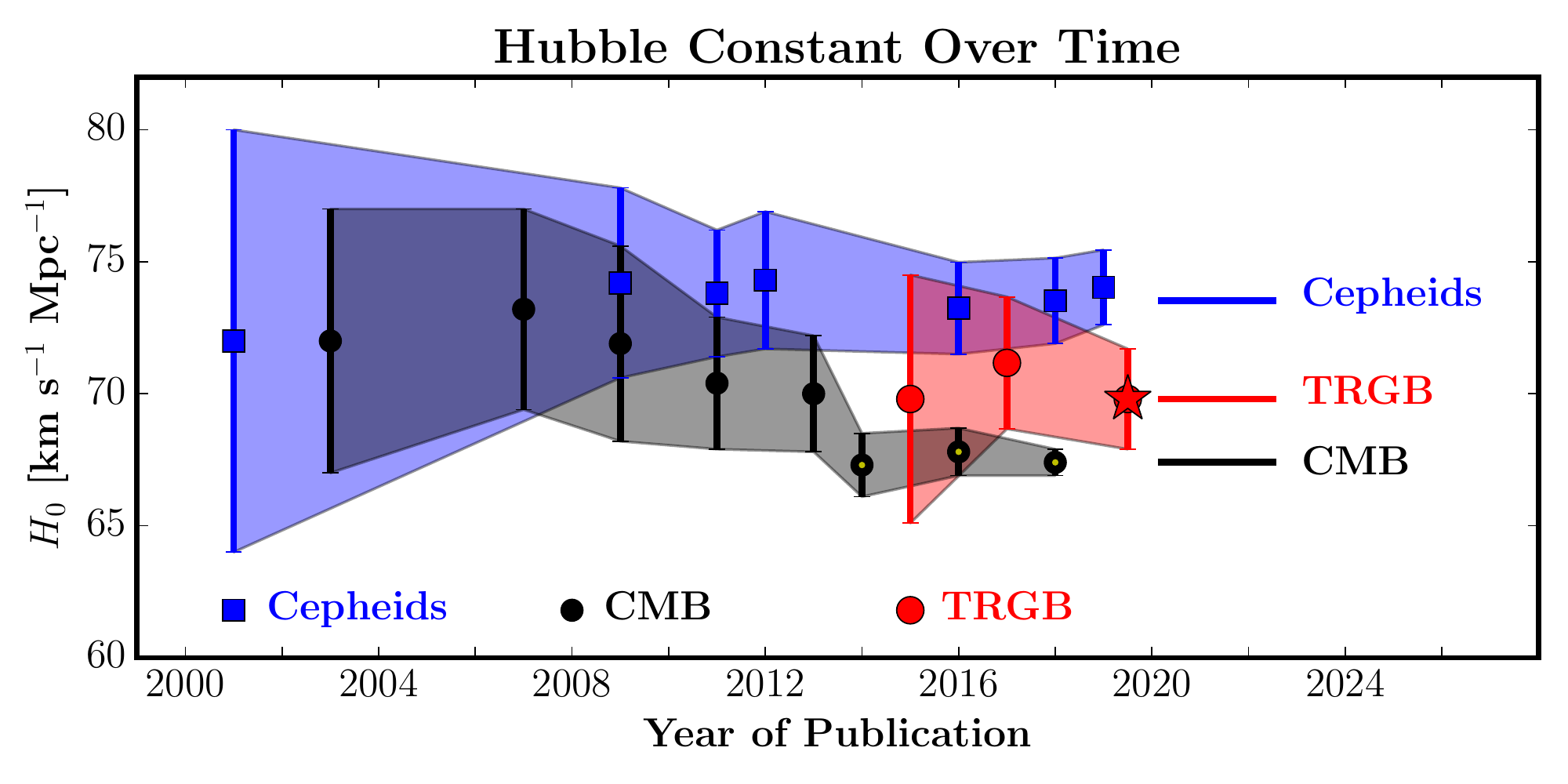}
 \caption{A plot of \ho values as a function of time. The points and shaded region in black are those determined from measurements of the CMB; those in blue are Cepheid calibrations of the local value of \ho; and the red points are TRGB calibrations. The red star is the best-fit value obtained in this paper. Error bars are 1\sig.}
 \label{fig:howithtime}
\end{figure*}

We show in Figure \ref{fig:howithtime} a comparison of local Cepheid (in blue) and TRGB (in red) determinations  of \ho, as well as values based on CMB measurements (in black), plotted as a function of year of publication. The value of \ho determined in this paper is denoted by a red star, and falls between the values defining the current \ho tension. It favors neither method, and equally can be used to argue for evidence that there is no tension (but ignoring the Cepheid results), or that, combining the TRGB and Cepheid results,  it provides low-level additional evidence that there is tension between the local and CMB values of \ho.

\section{The Future}
\label{sec:future}

In the next few years, a number of ongoing studies  will help to sharpen the  current debate over the early-universe and locally-determined values of \ho. We list five of them here: 

\begin{enumerate}
\item A major improvement to the parallax measurements from \gaia  is expected in 2022. At that time, accurate parallaxes ($<<1\%$) will become available for both Milky Way TRGB stars and Cepheids. In addition, they will be available for RR Lyrae stars. Although fainter than TRGB stars or Cepheids, RR Lyrae stars can provide a completely independent zero point for the nearest galaxies, allowing further testing for hidden systematics.

\item \hst will continue to allow measurement of distances to galaxies containing TRGB stars and Cepheids that are host to \sne, thereby increasing the numbers of SNIa-calibrating galaxies. Already, additional \hst time has been awarded for both programs in Cycle 26 (Proposal 15640; Freedman, PI and Proposal 15642; Riess, PI).

\item The launch of the James Webb Space Telescope ($JWST$) in 2021 will allow TRGB stars to be measured at infrared wavelengths where these stars are brighter than they are in the optical, thereby increasing the volume out to which TRGB distances can be measured, and increasing the numbers of \sn calibrating galaxies. Increasing the numbers of calibrators is particularly important. As the uncertainty in the zero point is decreased, the small number of calibrators (and their dispersion in absolute magnitudes) will become the largest uncertainty in the local determination of \ho. Unfortunately $JWST$ will not be capable of significantly extending the reach of the Cepheid distance scale for a number of reasons:  Cepheids are bluer stars, and their maximum variability (discovery potential) occurs at optical wavelengths.  $JWST$, optimized for the infrared, is diffraction limited at 2$\mu$m.  At larger distances, crowding of Cepheids by RGB and brighter AGB stars at redder wavelengths, combined with the smaller amplitudes in the infrared will severely limit their discovery and ultimate accuracy in \ho.

\item With Advanced LIGO and Virgo, the expected detection of significant numbers of  gravitational-wave events for neutron star -- neutron star coalescing binaries may provide a Hubble constant to 2\% accuracy within  5 years \citep{chen_fishbach_holz_2018}; see, however, \citet{shafieloo_keeley_linder_2018}, who note that the accuracy for this method in the near-term will still be dependent on the adoption of an underlying cosmological model.

\item The use of strong gravitational lens systems for measuring \ho  will provide a completely independent measure of \ho, and shows promise for  a 1\% determination of \ho in future years as hundreds, and possibly thousands of time-delay lens systems are discovered in future surveys \citep[e.g.,the H0LiCOW program][]{suyu_2017}.

\end{enumerate}

\pagebreak

\section{Summary }
\label{sec:summary}

The major result from this paper is  the construction and calibration of a new and  independent  distance scale for the local universe using the TRGB method, calibrating the absolute   distances to \sne in several independent surveys. We determine a value of the Hubble constant of \ho = 69.8 $\pm$ 0.8 ($\pm$1.1\% stat) $\pm$ 1.7 ($\pm$2.4\% sys) \hounits. This value differs only at the 1.2\sig level from the most recent \citet{planck_2018} inferred value of \ho.  It is smaller than  previous estimates of the Cepheid calibration of \sne \citep{freedman_2012, riess_2019}, but still agrees well at the 1.7\sig level. The TRGB method provides an opportunity to test for systematics in the Cepheid-based determination of \ho, which is significantly discrepant with that inferred from Planck. As we have demonstrated, the precision of the TRGB method is high, and future near-term improvements will continue to increase its accuracy.

\begin{figure*} 
\centering 
\includegraphics[width=1.0\textwidth]{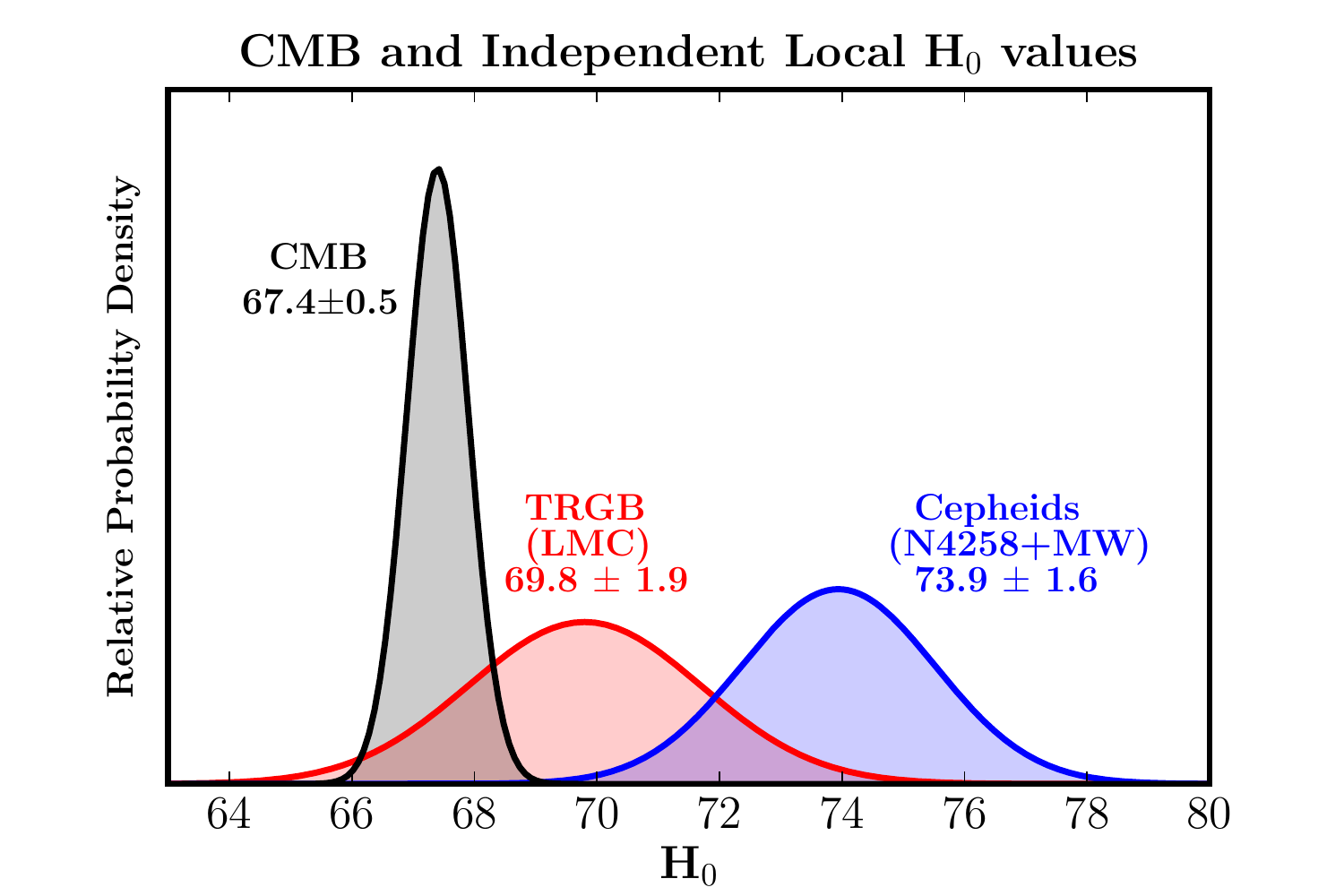}
\caption{Completely independent calibrations of \ho. Shown in red is the probability density function based on our LMC \cchp TRGB  calibration of \cspi \sne; in blue is the Cepheid calibration of \ho \citep{riess_2016}, using the Milky Way parallaxes and the maser distance to NGC~4258 as anchors (excluding the LMC).  The Planck value of \ho is shown in black.}
 \label{fig:HoMW4258pdfs}
\end{figure*}

In Figure \ref{fig:HoMW4258pdfs}, we compare the \ho probability density distributions for the TRGB, calibrated with the distance to the LMC; and Cepheids, calibrated with Milky Way parallax distances and the maser distance to NGC 4258  (and excluding the LMC calibration for Cepheids).  \citet{riess_2019} determine a value of \ho =  73.94 $\pm$ 1.58 \hounits based on the Milky Way and NGC 4258 calibration alone. This comparison is particularly important because the two determinations have minimal overlap in their systematics.  That is, the TRGB \ho value is calibrated via the distance to the LMC\footnote{We note that the result is unchanged (and the tension insignificantly so), whether or not we include the $Spitzer$-based Cepheid distance modulus to the LMC for the TRGB calibration.}, using the \cspi sample of \sne, whereas the Cepheid \ho value, in this particular comparison, is calibrated via the maser in NGC~4528 and Milky Way parallaxes, and uses the Supercal distant \sne sample. There are no measurements in common between these two methods, and therefore, the results are an excellent test of systematics. The independent TRGB and Cepheid results differ at the  $\pm$1.7\sig level. However, if one compares the astrophysical methods (by combining the two local measurements, TRGB and Cepheids), treating them as independent measurements,  a value of \ho = 72.26 $\pm$ 1.19 \hounits is found, resulting in a 3.7\sig tension with the Planck results.

We emphasize again that the methods of the TRGB and Cepheids are entirely independent. The  RGB stars are an old metal-poor halo population, whereas the Cepheids are a young metal-rich disk population. The physics of the helium flash for red giants and the pulsation mechanism for Cepheids are unrelated. The lines of sight to the supernova host galaxies have unrelated dust columns, one through the halo and the other through the gas-rich disk of the parent galaxy. Thus, the quantitative inter-comparison of these two methods provides an external test of the level at which  systematic errors are  independently affecting each of the two methods. For very nearby galaxies ($<$10 Mpc) the agreement is excellent: the scatter in the galaxy-to-galaxy comparison for the TRGB and Cepheid distances in common amounts to only $\pm$0.05 mag, or 2\% (see \S\ref{sec:trgbceph}).  The scatter in the galaxy-to-galaxy comparison for the TRGB and Cepheid distances for the \sn host galaxies alone is significantly larger, amounting  to $\pm$0.17 mag (\S\ref{sec:trgbceph}), and it is larger than what is expected from the published error bars.  The scatter in the local Hubble diagram (velocity versus distance) for the TRGB stars is 1.4 times lower than the scatter in the equivalent diagram for the Cepheids, indicating that the TRGB distances are more precise for these larger distances.

The TRGB method  has a number of advantages when applied to red giant branch stars in the halos of galaxies: these include low extinction by dust, low crowding/blending, and a metallicity effect that can be empirically calibrated directly for the TRGB stars themselves. In the $I$-band, there is almost no dependence on metallicity. There is also no need for multiple epochs of observations or concerns of different slopes with period, as in the case of Cepheid variables. In addition, the host masses of our TRGB host-galaxy sample are more massive, on average, than the galaxies in the Cepheid sample, thereby better matching  the range of masses of the \cspi distant sample. The largest systematics in the TRGB distance scale at present are: 1) the absolute zero point (currently set by the distance to the LMC), which determines the absolute magnitude of the TRGB,  for which we determine M$_I^{TRGB}$ = -4.05 $\pm$ 0.04 mag, and 2)  the small number of calibrating galaxies giving rise to the error on the mean for the  \sne calibration.

Ultimately, an unambiguous resolution of the \ho tension will require a local measurement of \ho to better than 1\%, a goal beyond the reach of current data sets. We note that the TRGB method holds considerable promise of improved accuracy in the near future as \gaia will provide a zero-point calibration to better than 1\% accuracy; and the number of calibrators can continue to be increased using both \hst and $JWST$, to bring down the uncertainty in the calibration of \sne to below the 1\% level.

We close by reiterating that the new TRGB results do not resolve the current \ho tension. Stated most simply, they agree with both the Planck and Cepheid \ho values. If taken alone, the TRGB results compared to those of Planck would suggest that there is no need for additional physics beyond the current standard cosmological model. However, there is strong motivation for using independent local measurements to test the standard model and its extrapolation to the present day. As noted above, if we combine the TRGB and Cepheid measurements and determine an independent local \ho value, the tension with Planck is at a 3.7\sig level, a significant tension, albeit lower than that seen for the Cepheids alone. Our results suggest that there is more work to be done to reduce systematic errors in the local distance scale before additional physics beyond the standard model is unequivocally called for. Whichever way the tension ultimately resolves, confirming the standard model, or pointing the way to additional physics, this issue remains one of the most important in cosmology today.

\acknowledgements

Support for program \#13691 was provided by NASA through a grant from the Space Telescope Science Institute, which is operated by the Association of Universities for Research in Astronomy, Inc., under NASA contract NASA 5-26555.
The \cspi has been supported by the National Science Foundation under grants AST0306969, AST0607438, AST1008343, AST1613426, and AST1613472.
M.G.L. is supported by a grant from the National Research Foundation (NRF) of Korea, funded by the Korean Government (NRF-2017R1A2B4004632).
Partial support for this work was provided by NASA through Hubble Fellowship grant \#51386.01  to R.L.B. by the Space Telescope Science Institute, which is operated by the Association of  Universities for Research in Astronomy, Inc., for NASA, under contract NAS 5-26555.
Computing resources used for this work were made possible by a grant from the Ahmanson Foundation.
This research has made use of the NASA/IPAC Extragalactic Database (NED), which is operated by the Jet Propulsion Laboratory, California Institute of Technology, under contract with the National Aeronautics and Space Administration.
Some of the data presented in this paper were obtained from the Mikulski Archive for Space Telescopes (MAST). STScI is operated by the Association of Universities for Research in Astronomy, Inc., under NASA contract NAS5-26555. 
We thank the {\it Observatories of the Carnegie Institution for
Science} and the {\it University of Chicago} for their support of our long-term research into the calibration and determination of the expansion rate of the Universe. 
W.L.F. thanks Stanford University for two visits during which time part of this paper was written. We thank Adam Riess for helpful correspondence concerning the LMC calibration. We also thank the anonymous referee for constructive comments.

{\it Facilities:} $HST$ ($ACS$)


\appendix 

\section{Estimate of Reddening to the LMC TRGB Calibration Sample}
\label{App:appendix_lmcred}

For completeness, we describe briefly the method used by \citet{madore_freedman_2019b} for determining the reddening to the LMC outer-field TRGB stars.

This method employs multi-wavelength observations in a similar manner to that which has been widely in use for determining distances and reddenings to Cepheids
in nearby galaxies. Measurements of the apparent magnitudes of the TRGB at multiple
wavelengths are used to derive individual apparent moduli as a function of inverse wavelength;  a reddening law can then be fit to these data, giving
the total line-of-sight extinction. These methods were first introduced by \citet{freedman_grieve_madore_1985,freedman_1988} with recent examples being \citet{scowcroft_2013}, who used  9-band photometry to study the Cepheids in \ic, and \citet{rich_2014} who used 7-band photometry to determine the distances and reddenings to  Cepheids in \ngc 6822.

In this context, \citet{madore_freedman_2019b} use the two low-reddening galaxies, \ic and the SMC, as {\it differential} calibrators relative to the LMC. Both of these galaxies also have measured apparent magnitudes  for their TRGB populations at the same set of wavelengths as the LMC sample.  Differential distance moduli and differential reddenings can then been obtained with high precision between \ic  and the SMC, as well as between them individually and the LMC.

The zero point of the reddening calibration is set using the (low) value of the total line-of-sight extinction of $E(B-V) = $ 0.02 $\pm$ 0.003~mag for \ic, which was derived from the all-sky reddening maps published by \citet{schlafly_finkbeiner_2011}. The zero point of the reddening-corrected, absolute magnitude calibration is in turn set at the LMC itself, where the geometric distance modulus of 18.477 $\pm$ 0.026~mag has been recently determined by \citet{pietrzynski_2019}. The multi-wavelength TRGB absolute magnitude calibrations can now be used to determine both distances and
reddenings for any future TRGB observations made at three or more wavelengths.

For completeness these calibrations are given here below:

$$M_V = -2.44 + 1.00 \times [(V-I)_o - 1.6]$$

$$M_I = -4.04 $$

$$M_J = -5.16 - 0.85 \times [(J-K)_o - 1.0]$$

$$M_H = -5.93 - 1.62 \times [(J-K)_o - 1.0]$$

$$M_K = -6.12 - 1.85 \times [(J-K)_o - 1.0]$$

\begin{figure*}
\figurenum{A1}
\centering
\includegraphics[width=1.0\textwidth]{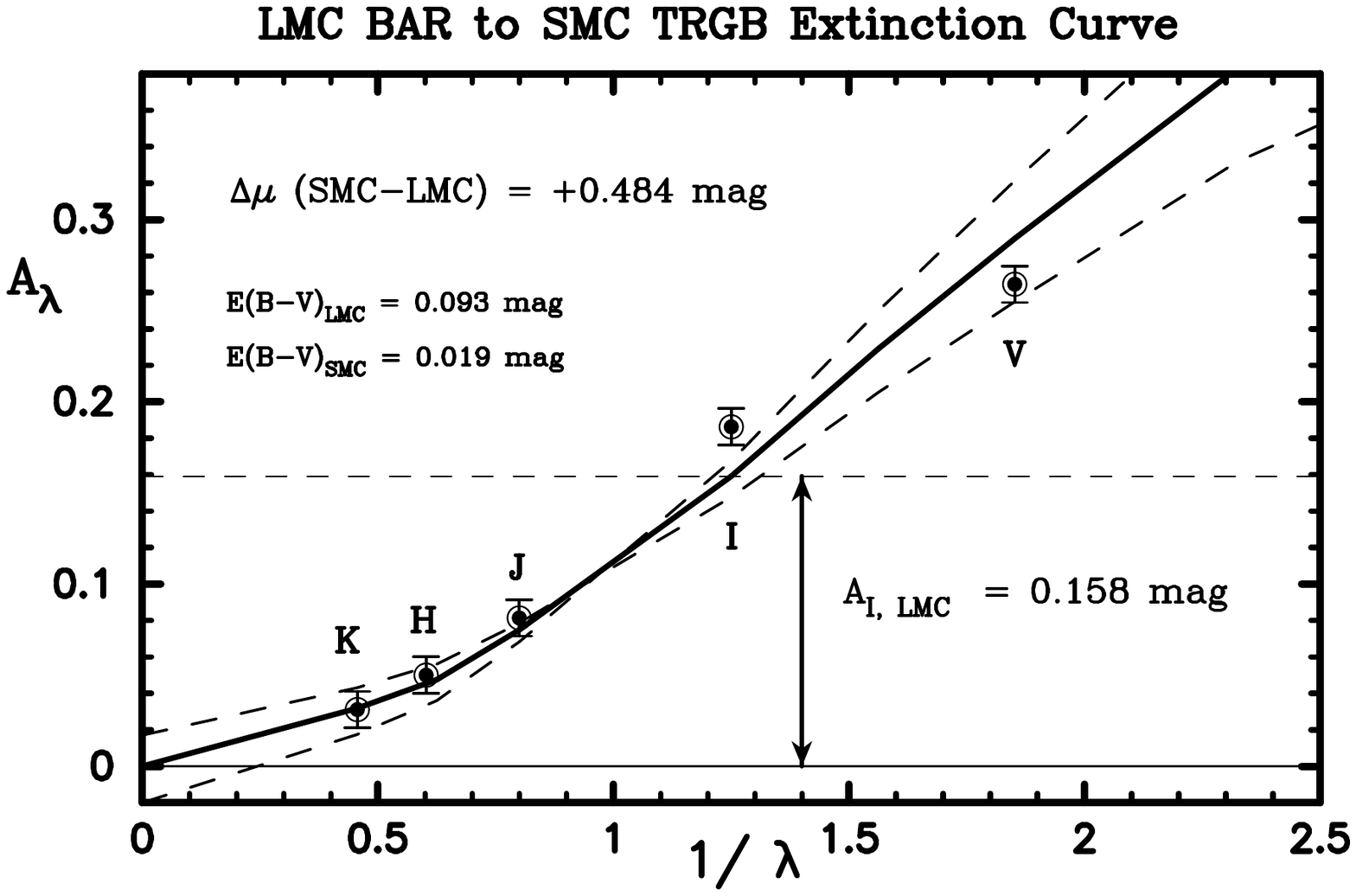}
\caption{Multi-wavelength extinction as a function of inverse wavelength for TRGB stars in the bar of the LMC. Differencing the apparent magnitudes of the TRGB stars in the LMC with respect to those in the SMC, measured at five different wavelengths from the near infrared ($JHK$) to the optical ($VI$), allows one to simultaneously solve for the difference in true distance modulus $\Delta\mu =$  +0.484 mag and derive the total line-of-sight reddening to the LMC TRGB stars. In this case, a value of  $E(B-V) = $ +0.093~mag is obtained.  The flanking dashed lines shown are $\pm$2\sig about the fit.  The horizontal dashed line shows the value of A$_I$.
\label{fig:fa1}}
\end{figure*}

\begin{figure*}
\figurenum{A2}
\centering
\includegraphics[width=1.0\textwidth]{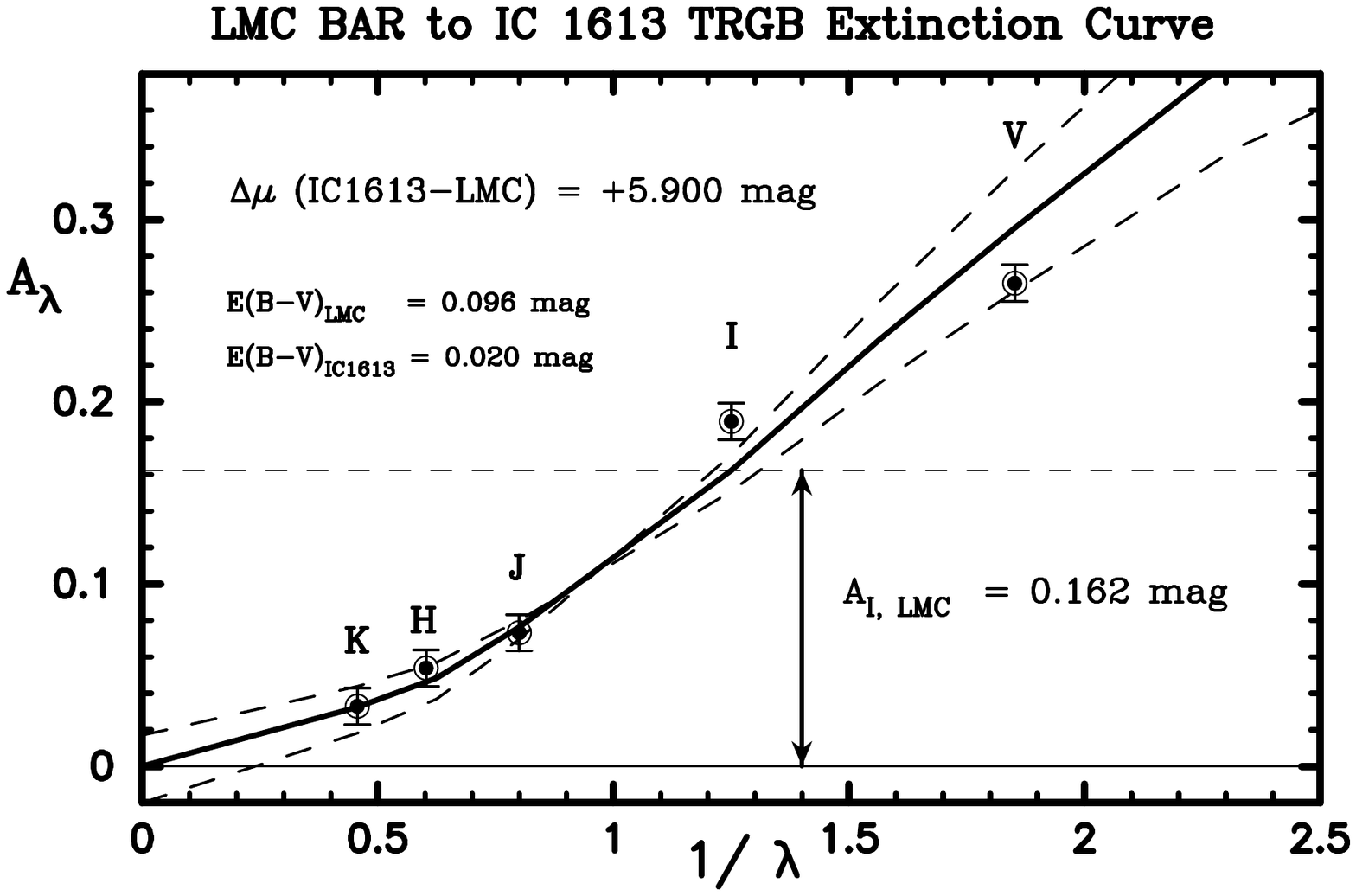}
\caption{Multi-wavelength extinction as a function of inverse wavelength for TRGB stars in the bar of the LMC. Differencing the apparent magnitudes of the TRGB stars in the LMC with respect to those in the dwarf galaxy IC~1613, measured at five different wavelengths from the near infrared 
($JHK$) to the optical ($VI$), allows one to simultaneously solve for the difference in true distance modulus $\Delta\mu =$  +5.900~mag and derive the total line-of-sight reddening to the LMC TRGB stars. In this case, a value of  $E(B-V) = $ +0.096~mag is obtained.  Dashed lines are as in Figure \ref{fig:fa1}.
\label{fig:fa2}}
\end{figure*}

\clearpage

\section{Flow- and Group-Corrected Velocities for Nearby Calibrators}
\label{App:appendix_NEDvel}

The velocities for the nearby calibrators have been corrected using the linear multi-attractor model available in NED. We provide additional notes below for individual galaxies.  Further corrections are made if the galaxy is a member of a group, as noted below.

\noindent
NGC 3021  is a member of the triple system KTG 26. The averaged heliocentric velocity of the
          triple is +1492 km/sec; and the flow correction applied was +394 km/s. 

\noindent
NGC 3370  is the brightest member of a small group. The flow-corrected velocity as listed was averaged over 24 group members.

\noindent
NGC 4038/39  is an interacting double. The averaged velocity of the two components from NED was used.
     
\noindent
NGC 5584  is a member of the Virgo III association.  The averaged heliocentric velocities of thirteen group members is +1608 km/sec; and the flow correction applied was +375 km/s. 
          
\noindent
M101  is an isolated galaxy whose NED flow-corrected velocity is +455 km/s.

\noindent
NGC 1309  is a member of a small group whose mean flow-corrected velocity is +1864 km/s.

\noindent
NGC 1365  is a member of the Fornax cluster whose local expansion velocity is taken to be +1306 km/sec \citep{madore_1999}.

\noindent
NGC 1448  is a member of a triplet which has a mean heliocentric velocity is 1142 km/sec and a flow correction of -95 km/sec.

\noindent
NGC 4424  is a member of the Virgo Cluster, whose local expansion velocity is taken to be +1050 km/s.

\noindent
NGC 4526  is a member of the Virgo Cluster, whose local expansion velocity is taken to be +1050 km/s.

\noindent
NGC 4536  is a member of the Virgo Cluster, whose local expansion velocity is taken to be +1050 km/s.

\noindent
NGC 3627 (M66) is a member of the Leo Group, whose local expansion velocity is taken to be +689 km/s.
\medskip

\noindent
NGC 3368 (M96) is a member of the Leo Group, whose local expansion velocity is taken to be +689 km/s.
\medskip

\noindent
NGC 1316  is a member of the Fornax cluster whose local expansion velocity is taken to be +1306 km/sec \citep{madore_1999}.

\noindent
NGC 1404  is a member of the Fornax cluster whose local expansion velocity is taken to be +1306 km/sec  \citep{madore_1999}.

\noindent
NGC 1015  is isolated.

\noindent
NGC 2442  is a member of the NGC 2442 Group. Group velocity from NED.

\noindent
NGC 3447  is a member of the M96 Group whose mean heliocentric velocity is 1199 km/s, with a calculated flow correction of +248 km/s.

\noindent
NGC 3972  is a member of the Ursa Minor Group \citep{fouque_1992}.

\noindent
NGC 3982  is a member of the Ursa Minor Group \citep{fouque_1992}.

\noindent
NGC 4639  is a member of the Virgo Cluster, whose local expansion velocity is taken to be +1050 km/s.

\noindent
NGC 5917  is a member of an interacting pair whose mean heliocentric velocity is +1921 km/sec with a calculated flow correction of +323 km/sec.

\noindent
NGC 7250  is isolated. 

\noindent
UGC 09391  is isolated.

\clearpage


\end{document}